\documentclass[twocolumn]{aastex631}
\usepackage[version=4]{mhchem}

\newcommand{\swampe}{\texttt{SWAMPE} }
\newcommand{\revise}[1]{{#1}}

\shorttitle{Asymmetry and Dynamical Constraints in 2-Limbs Retrieval for WASP-39 b}
\shortauthors{Zixin Chen et al.}

\graphicspath{{./}{figs/}}

\begin{document}
\title{Asymmetry and Dynamical Constraints in 2-Limbs Retrieval of WASP-39 b Inferring from JWST Data}

\correspondingauthor{Jianghui Ji}
\email{jijh@pmo.ac.cn}

\author{Zixin Chen}
\affiliation{CAS Key Laboratory of Planetary Sciences, Purple Mountain Observatory, Chinese Academy of Sciences, Nanjing 210023, People's Republic of China}
\affiliation{School of Astronomy and Space Science, University of Science and Technology of China, Hefei 230026, People's Republic of China}

\author[0000-0002-9260-1537]{Jianghui Ji}
\affiliation{CAS Key Laboratory of Planetary Sciences, Purple Mountain Observatory, Chinese Academy of Sciences, Nanjing 210023, People's Republic of China}
\affiliation{School of Astronomy and Space Science, University of Science and Technology of China, Hefei 230026, People's Republic of China}
\affiliation{CAS Center for Excellence in Comparative Planetology, Hefei 230026, People's Republic of China}

\author[0000-0003-0740-5433]{Guo Chen}
\affiliation{CAS Key Laboratory of Planetary Sciences, Purple Mountain Observatory, Chinese Academy of Sciences, Nanjing 210023, People's Republic of China}
\affiliation{CAS Center for Excellence in Comparative Planetology, Hefei 230026, People's Republic of China}

\author[0000-0001-9585-9034]{Fei Yan}
\affiliation{Department of Astronomy, University of Science and Technology of China, Hefei 230026, People's Republic of China}

\author[0000-0003-2278-6932]{Xianyu Tan}
\affiliation{Tsung-Dao Lee Institute, Shanghai Jiao Tong University, 520 Shengrong Road, Shanghai 200127, People's Republic of China}
\affiliation{School of Physics and Astronomy, Shanghai Jiao Tong University, 800 Dongchuan Road, Shanghai 200240, People's Republic of China}

\begin{abstract}

Transmission spectroscopy has provided unprecedented insight into the makeup of exoplanet atmospheres. A transmission spectrum contains contributions from a planet's morning and evening limbs, which can differ in temperature, composition and aerosol properties due to atmospheric circulation. While high-resolution ground-based observations have identified limb asymmetry in several ultra-hot/hot exoplanets, space-based studies of limb asymmetry are still in their early stages. The prevalence of limb asymmetry across a broad range of exoplanets remains largely unexplored.
We conduct a comparative analysis of retrievals on transmission spectra, including traditional 1D approaches and four 2D models that account for limb asymmetry. Two of these 2D models include our newly proposed dynamical constraints derived from shallow-water simulations to provide physically-motivated temperature differences between limbs. Our analysis of WASP-39 b using JWST observations and previous combined datasets (HST, VLT, and Spitzer) strongly favors 2D retrievals over traditional 1D approaches, confirming significant limb asymmetry in this hot Jupiter. Within our 2D framework, unconstrained models recover larger temperature contrasts than dynamically-constrained models, with improved fits to specific spectral features, although Bayesian evidence cannot definitively distinguish between these 2D approaches.
Our results support the presence of homogeneous C/O in both the morning and evening atmospheres, but with temperature differences leading to variations in clouds and hazes. Using this treatment, we can study a larger sample of hot Jupiters to gain insights into atmospheric limb asymmetries on these planets.

\end{abstract}


\keywords{
Unified Astronomy Thesaurus concepts: Exoplanets (498); Exoplanet astronomy (486); Transmission spectroscopy (2133); Exoplanet atmospheres (487); Exoplanet atmospheric composition (2021)}

\section{Introduction} \label{sec:intro}

Transmission spectroscopy is a valuable tool for gaining insights into exoplanet atmospheres. A transmission spectrum describes the absorption \revise{and scattering }of photons from a star as they pass through the planet's atmosphere in the terminator region. This portion of the atmosphere is often assumed to be homogeneous, \revise{with uniform physical and chemical properties at any given pressure levels across different longitudes and latitudes of the terminator.} Using this simple yet effective one-dimensional (1D) assumption, significant breakthroughs in understanding exoplanet atmospheric composition have been made through observations over the past two decades \citep{Charbonneau2002,Vidal-Madjar2003,Sing2016}.

However, the transmission region is composed of both the morning and evening limbs, which may differ in temperature and composition, as suggested by 3D circulation models of hot giants \citep{Showman2002,Cooper2005,Showman2009,Menou2009,Rauscher2012,Mayne2013,Perez-Becker2013,
Amundsen2016,Mayne2017,Drummond2020}. Hot giants are believed to be tidally locked, with one side permanently facing their star, leading to a strong day-night contrast due to stellar radiation. In addition to this initial heat distribution, the dynamical circulation-driven redistribution, including east-west(morning and evening) differences, breaks the terminator area's homogeneity assumption.

Research about atmosphere dynamics has predicted that there would be a superrotation equatorial jet on a tidally locked hot giant, which transports and significantly redistributes energy and heat \citep{Perez-Becker2013,2021MNRAS.501...78P,2021ApJ...908..101R}.
The evening limb is downstream of the heat flow, while the morning limb is upstream\revise{ \citep{2007Natur.447..183K,2014Sci...346..838S,2024NatAs...8..879B}}. Meanwhile, the hotspot on the hot giant is shifted to the east from the substellar point, making the hot region closer to the evening limb than the morning one.
Therefore, temperatures are expected to be higher in the evening than in the morning, and the atmospheric composition will more closely resemble that of the dayside. This contrasts with the morning limb, which retains more residual products from the nightside atmospheric processes.

Several studies have combined 3D modelling results with observational effect \citep{Fortney2010,Dobbs-Dixon2012}. Others focus on the influence of inhomogeneous clouds  \citep{Line2016,Kempton2017,Powell2019}. Currently,  \citet{Roth2024} conducted a large grid of non-grey global circulation models on highly irradiated hot gas giants, exploring the a-prior scatter in the population created by the different responses of atmospheric circulation to planetary parameters. They find the temperature contrast between the limbs due to the combined effect of phase shift and hot redistribution and predict the largest limb-to-limb temperature differences for planets with $T_{eq} \approx 2000~\mathrm{K}$.

While high-resolution ground-based observations have identified limb asymmetry in several gas giants \citep{Ehrenreich2020,Cont2024,Nortmann2024,YangYang2024}, space-based studies of limb asymmetry in gas giants are still in their early stages.

To make the best use of JWST's strengths, there are approaches to mimic the transit curve with a non-circle plane to consider the inhomogeneous atmosphere and even get the individual transmission spectra of each limb instead of an averaged spectrum from the simplest circle assumption \citep{Espinoza2021,Jones2022}.
To date, the evidence of limb asymmetry has been seen in two gas giant's transmission spectra, WASP-107 b ($T_{eq} = 770\,\mathrm{K}$)  \citep{Murphy2024} and WASP-39 b ($T_{eq} =1200\,\mathrm{K}$) \citep{Espinoza2024}. Both are quite tempered compared to those ultra-hot giants ($>2000\,\mathrm{K}$) regarded to have significant limb-to-limb temperature contrast.

Despite extracting individual spectra, the traditional spectral interpretation using averaged spectra will remain the dominant approach for the foreseeable future due to limited observational resources.
Our objective is to improve the retrieval of averaged transmission spectra by incorporating dynamic atmospheric knowledge into the process.
Atmospheric temperature is a key factor in transmission spectrum models, as it directly influences scale height and the composition derived from equilibrium chemistry. We propose that it is possible to estimate the temperature difference at the planetary limb using a shallow water model and incorporate this as an additional dynamical constraint to account for the morning and evening limb temperature differences during retrieval.

We choose WASP-39 b as a demonstration case for our method.
WASP-39 b is a hot Saturn with a mass of $0.28\,M_\text{Jup}$  and its radius  $R=1.27\,R_\text{Jup}$ \citep{faedi2011wasp}. This gas-giant orbits around its star in 4.05528 days which implies that it is likely to be tidally locked. The very short period and low-density features made WASP-39 b a good target for transmission spectroscopic observation.
Transmission spectra obtained by ground-based \citep{Nikolov2016} and space-based telescopes \citep{fischer2016hst,Sing2016,Wakeford2017} have been interpreted using different models assumption of thermal structure, chemical equilibrium and aerosol treatments \citep{tsiaras2018population,pinhas2018retrieval,pinhas2019h2o,kirk2019lrg,welbanks2019degeneracies,Kawashima2021}, but have failed to reach relatively consistent conclusions regarding water abundance and metallicity.
Retrieval analysis accounting for stellar activity shows that stellar activity plays a negligible role in the transmission spectrum of WASP-39 b \citep{kirk2019lrg}.
The JWST Transiting Exoplanet Community Early Release Science Program (ERS Program 1366) applied all three instruments in near-IR using four observation modes \citep{alderson2023early,Ahrer2023a,feinstein2023early, Rustamkulov2023}. And a following observation was made by MIRI to further confirm the \ce{SO2} absorption \revise{\citep{2024Natur.626..979P}}.
JWST has detected various molecular species, including \ce{CO2}, \ce{H2O}, and \ce{CO} \citep{alderson2023early,Ahrer2023a,feinstein2023early, Rustamkulov2023}, as well as alkali metals such as sodium \citep{Rustamkulov2023} and potassium \citep{feinstein2023early}. Additionally, a tentative detection of \ce{SO2} \revise{\citep{Rustamkulov2023, 2024Natur.626..979P}}, a photochemical byproduct, indicates active atmospheric \revise{photo}chemistry \citep{Tsai2023c}.

In this work, we employed a dynamical module to precompute a fixed morning-to-evening temperature difference, which was subsequently used as input for the 2D retrieval framework. This approach enhances the accuracy of temperature characterization and mitigates the overestimation of temperature differences that can arise in unconstrained retrievals. Specifically, we applied our retrieval method to WASP-39 b using the latest JWST data along with previous combined datasets.
In Section~\ref{sec:method}, we provide a detailed description of our retrieval method. Section~\ref{sec:apply} outlines the application of empirical observations to our retrieval methodology. In Section~\ref{sec:discuss}, we present a comparative analysis of the results for WASP-39 b across different models and datasets. This analysis focuses on three key aspects: temperature , atmospheric composition (metallicity and C/O ratio), and aerosols (clouds and hazes) on each limb.
In final, we summarize our results and key lessons, where we discuss the limitation and the prospects for future investigation of inhomogeneous limbs in exoplanetary transmission spectra.

\section{Method}\label{sec:method}

\subsection{Calculate 1D Radiative Transfer with \texttt{PLATON}}
We configure \texttt{PLATON} (version 5.3) \citep{Zhang2019,Zhang2020a} to implement forward modeling in the subsequent retrieval analyses, which conducts 1D radiative transfer to calculate the transmission spectrum of a plane-parallel atmosphere with hydrostatic equilibrium.
The atmosphere is assumed to have an isothermal temperature of $T$, which is divided into 50 layers, Log-equally spaced from $10^3$ to $10^{-9}$ bar, with the reference planet radius ($R_\text{p,1bar}$) set at 1 bar. The gas absorption, collisional absorption (\ce{H2-H2} and \ce{H2-He}) \citep{Richard2012,Karman2019}, and scattering absorption are taken into account. The gas abundances are obtained from the equilibrium chemistry abundance grid pre-calculated by \texttt{GGChem}  \citep{Woitke2018}, which is a function of species name, temperature, pressure, metallicity (Z), and C/O ratio. A total of \revise{30} atomic and molecular species are considered, including: \ce{H}, \ce{He}, \ce{C}, \ce{N}, \ce{O}, \ce{Na}, \revise{\ce{Fe}, \ce{Ca}, \ce{Ti}, \ce{Ni}, }\ce{K}, \ce{H2}, \ce{H2O}, \ce{CH4}, \ce{CO}, \ce{CO2}, \ce{NH3}, \ce{N2}, \ce{O2}, \ce{O3}, \ce{NO}, \ce{NO2}, \ce{H2S}, \ce{HCN},\ce{OH}, \ce{PH3}, \ce{SiO}, \ce{SO2}, \ce{TiO}, and \ce{VO} \citep{Allard2016,Tennyson2018TheEA,Allard2019}.
The radiative transfer modeling is based on the opacity line list with a resolution of $\lambda /\Delta \lambda = 10 000$ as suggested by \citet{Zhang2019}.
The clouds are described as an optically thick cloud deck with a cloud-top pressure of $P_\text{c}$. The hazes are parameterized to have a Rayleigh-like scattering with a slope of $\gamma$ at an amplitude of $A$. We prescribe uniform or log-uniform priors for all free parameters (see Table \ref{tab:platon-para}).

As a default, \texttt{PLATON} assumes an isothermal 1D atmosphere with uniform clouds (hereafter 1D cloudy). The forward model consists of seven basic free parameters: $R_\text{p,1bar}$, $\log Z$, $T$, C/O, $\log P_\text{c}$, $\gamma$ and $\log A$.

\begin{deluxetable}{lll}[htb] \label{tab:platon-para}
  \tablecaption{Free parameters of \texttt{PLATON} used in this work}
  \tablehead{Parameters & Description(unit)                                      & Prior  }
  \startdata
  $R_\mathrm{p,1bar}$        & Reference radius at 1 bar \revise{($R_\text{Jup}$)}   & $\mathcal{N} [1.28,0.12]$  \\
  $T$     & Limbs temperature(K)    & $\mathcal{U}$[700,1500]   \\
  $\log Z$              & Log metallicity(solar)  & $\mathcal{U} $[-1,3]       \\
  C/O                 & Carbon-to-oxygen ratio     & $\mathcal{U} $[0.05,2]     \\
  $\log P_\mathrm{c}$& Log cloud-top pressure(bar)               & $\mathcal{U} $[-6,2]       \\
  $\gamma$              & Rayleigh-like scattering slope                & $\mathcal{U} $[-20,2]      \\
  $\log A$              & Log scattering enhancement factor     & $\mathcal{U} $[-4,12]      \\
  $\Phi$                & Patchy cloud cover factor                     & $\mathcal{U} $[0,1]        \\
  \enddata
\end{deluxetable}

\subsection{Linear combination of 1D models}

Accounting for atmospheric inhomogeneity in the transmission region  such as variations in temperature structure, chemistry, and cloud coverage due to atmospheric circulation predicted by GCMs can result in distinct spectral features in transmission spectra \citep{Fortney2010}. To address this, several studies have explored the use of linear combinations of multiple 1D atmospheric models \revise{\citep{Line2016,Kempton2017,2017MNRAS.469.1979M,Welbanks2021,Welbanks2022,Li2023}}.

A straightforward improvement for considering limb inhomogeneity involves considering cloud coverage, introducing an additional free parameter, the cloud coverage factor $\phi$, into the 1D cloudy model. This approach, referred to as the 1D patchy cloudy model, represents the planetary atmosphere as a linear combination of a clear sector and a cloudy sector. The cloudy sector is described by the seven free parameters of the 1D cloudy model, while the clear sector shares the same parameter values except for $\log P_\text{c}$, which is set to infinity to indicate no clouds. Consequently, the 1D patchy cloudy model includes a total of eight free parameters: $R_\text{p,1bar}$, $\log Z$, $T$, C/O, $\log P_\text{c}$, $\gamma$, $\log A$ and $\phi$.

The 2D models employ similar configurations to describe the morning and evening limbs. In contrast to the 1D patchy cloudy model, the 2D models consist of two equal-weight sectors representing the two limbs, enabling a more detailed parameterization of each limb's properties. Parameters with the subscripts 'morn' and 'even' refer to the morning and evening limbs, respectively, while parameters without subscripts are assumed to be identical for both limbs. Here we configure four 2D models to investigate the impact of fixing the temperature difference and reducing the number of parameters related to composition and aerosol properties.

In the temperature retrieval process, a formal approach involves allowing two temperature parameters, $T_\text{morn}$ and $T_\text{even}$, to vary independently. This approach is referred to as the '2Dfree' method. \revise{For the alternative '2Dfixed' method, we adopt a predefined temperature difference $\Delta T = 150~$K between morning and evening derived from the SWAMPE model (see Section 3.2 for justification). In this case,} only $T_\text{morn}$ is treated as a free parameter. The evening temperature, $T_\text{even}$, is then computed automatically as $T_\text{even} = T_\text{morn} + \Delta T$.

Beyond temperature, we also investigate whether the assumption of homogeneous composition introduces biases in retrievals. \revise{GCMs predicted that for planet with equilibrium temperature below 1800 K, global mixing homogenizes mean molecular weight, but intermittent cloud coverage can drive localized C/O variations \citep{2023A&A...671A.122H}. Therefore, the metallicity is assumed to be the same for both limbs across all 2D models, and C/O ratio is allow for test. }In Model A, the C/O ratios of the morning and evening limbs are assumed to be identical. In contrast, Model B allows for variability, treating the C/O ratios of the morning and evening limbs as independent parameters, denoted as C/O$_\text{morn}$ and C/O$_\text{even}$, respectively.

\subsection{Calculate Temperature Difference with \swampe}\label{sec:swampe}

The addition of dynamical module represents a vital innovation and modification to the traditional 1D approach and the original \texttt{PLATON} code. This module aims to simulate global atmospheric circulation on the planet, with a primary objective of estimating the temperature difference $\Delta T$ between the morning and evening terminators resulting from atmospheric dynamics.

Previous studies have explored atmospheric circulation using high-complexity 3D atmospheric circulation models, particularly those addressing the diverse chemical and physical processes across the parameter space of hot Jupiters \citep{Roth2024}. While such models provide valuable insights, they are computationally expensive and time-intensive, making them impractical for retrieval studies, which typically require the computation of thousands of forward models. Conversely, simpler 1D models often fail to capture the inherently 3D nature of atmospheric processes, highlighting the need for intermediate approaches that balance complexity and computational efficiency\revise{ \citep{2022ApJ...929...20M}}.

For these reasons, a shallow water model is employed in this work to study planetary atmospheric circulation. While more simplified than 3D models, 2D shallow water models effectively capture key dynamical processes. These models simulate the behavior of a thin fluid layer of constant density and variable thickness, governed by the shallow water equations.
These equations constitute a coupled system that governs the conservation of horizontal momentum and mass. They control the evolution of the longitudinal and meridional velocity components, as well as the thickness of the layer. Typically, the equations are solved numerically across longitude, latitude, and time.

We adopt \swampe, a Python package designed for modeling exoplanetary atmospheric dynamics \citep{Landgren2022}, to perform the shallow water model calculations. \citet{Landgren2023} demonstrated the use of \swampe to simulate global atmospheric circulation in sub-Neptune exoplanets, focusing on the effects of planetary rotation rates and radiative timescales across diverse stellar insolation regimes.

In our \swampe calculations, we adopted the system parameters of WASP-39 b, which has a planetary radius of $a = 1.27\,~R_\mathrm{J}$. Assuming the planet is tidally locked, its rotation period equals its orbital period, $\Omega = 1.793 \times 10^{-5}\,\text{rad/s}$. The reference geopotential height is defined as $\Phi = gH$, where $g = 4.07 \, \text{m}/\text{s}^2$ and $H = 1037 \, \text{km}$. Computational parameters include a time step $\text{d}t = 30 \, \text{s}$ and  \revise{a common spectral truncation used in global climate models, 42-wave triangular truncation (T42)}, corresponding to a spatial resolution of $128 \times 64$ in longitude and latitude, with a grid cell size of $2.81 ^{\circ}$.

To represent the strong irradiation contrast between the dayside and nightside of this hot Saturn-like planet, we set the relative radiative forcing amplitude to $\Delta\Phi / \bar{\Phi} = 1$, following \citet{Perez-Becker2013}. Both the radiative timescale $\tau_\mathrm{rad}$ and the advective timescale $\tau_\mathrm{adv}$ are set to be 1 day.

\begin{figure}[htb]
  \centering
  \includegraphics[width=\columnwidth, angle=0]{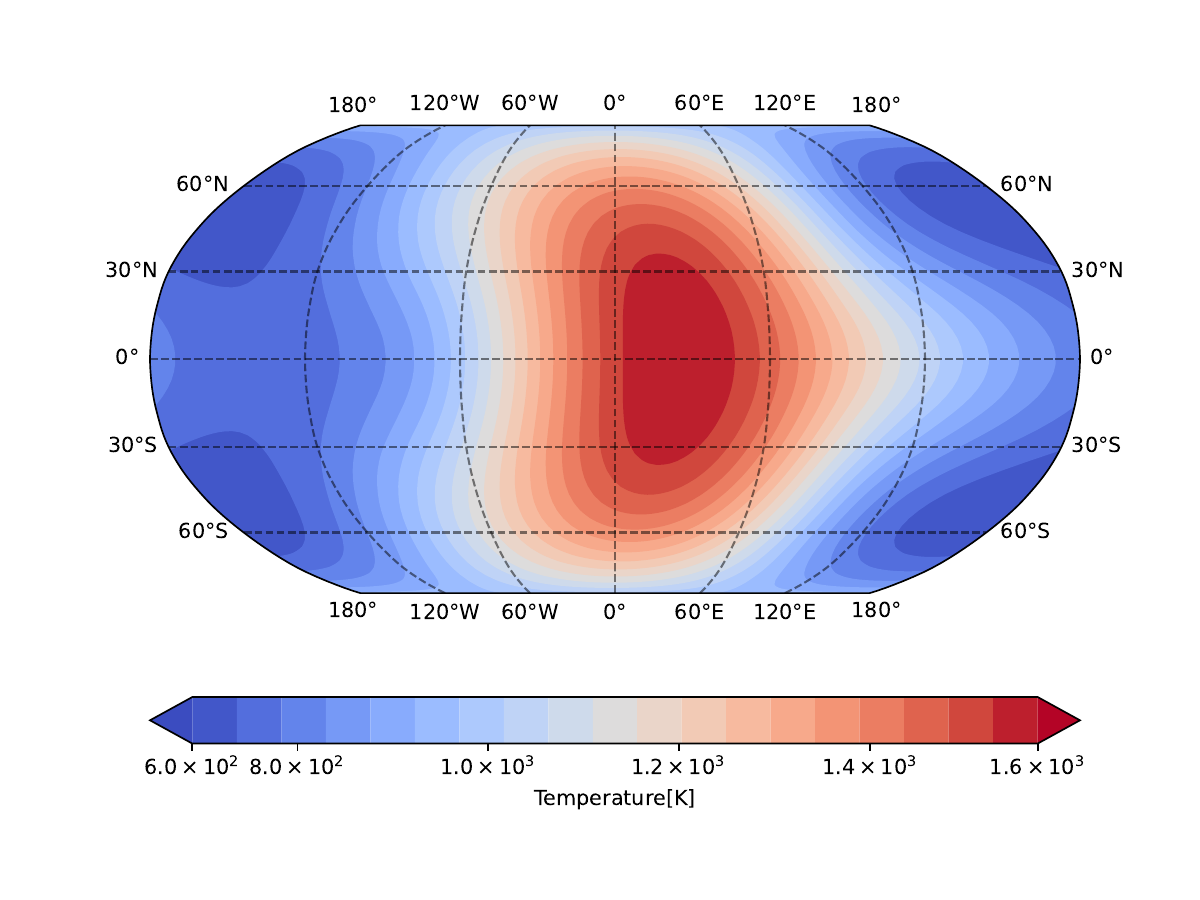}
  \caption{Temperature mapping from shallow water simulation of WASP-39 b.}
  \label{fig:temp_map}
\end{figure}

Fig~\ref{fig:temp_map} shows the simulated atmospheric temperature pattern of WASP-39 b. Using the geopotential height distribution generated by \swampe, we derive the corresponding temperature \revise{spatial }distribution as a preparatory step for subsequent analyses. The underlying physics of this mapping relies on the relationship between atmospheric heating and geopotential changes: heating increases the geopotential $\Phi$, while cooling decreases it. Thus, qualitative variations in $\Phi$ can be interpreted as changes in temperature, with a thicker upper atmospheric layer corresponding to higher temperatures.
We apply the idealized relationship $\Phi \approx RT$, where $R$ is the specific gas constant, to relate the geopotential height to temperature \citep{Holton1992,Landgren2023}. For hot Jupiters, the mean molecular weight of the atmosphere is approximately 2.4 times the mass of a hydrogen atom, corresponding to a $R$ value of $3464~\mathrm{J} \, \mathrm{kg}^{-1} \, \mathrm{K}^{-1}$. \revise{We estimated an appropriate opening angle of approximately 40 degrees based on \citet{2022MNRAS.510..620W}. This corresponds to 14 longitude cells in our SWAMPE grid (covering 39.4 degrees) centered on each terminator. The average temperature in each region was calculated as the mean of all grid points within this angular range.}
By subtracting the average temperature of the morning limb \revise{ (1110 K)} from that of the evening limb\revise{ (1260 K) }, we calculate a temperature difference of 150 K between the terminators in the case of WASP-39 b.

\subsection{Retrieval algorithm and model evidence}

The atmospheric retrievals are performed using a nested sampling algorithm, assuming uniform or log-uniform priors for all free parameters. The total log-likelihood is computed as the sum of the log-likelihood contributions from each observation. We use \texttt{PyMultiNest} \citep{Buchner2014} for the retrievals, setting 250 live points and a sampling efficiency of 0.8.

For each retrieval model, the marginalised likelihood (also referred to as the Bayesian evidence) is calculated. The ratio of Bayesian evidences between two models, known as the 'Bayes factor' B, provides a quantitative means of model comparison. It helps identify the model that best balances explanatory power with complexity \citep{Trotta2008}.
We follow the classification criteria outlined by \citet{Trotta2008}: when $|\ln B| < 1$, both models are considered to explain the data equally well. For $|\ln B| \geq 2.5$, the model with higher Bayesian evidence is moderately favored, while for $|\ln B| \geq 5$, it is strongly favored.

\section{Retrievals with Existing Observations}\label{sec:apply}

WASP-39 b was previously observed using the Space Telescope Imaging Spectrograph (STIS) onboard Hubble,  which covered the wavelength range of $0.29 - 1.025\mu\mathrm{m}$ \citep{fischer2016hst}. Later, the other instrument on Hubble, the Wide Field  Camera 3 (WFC3) was used to make defined detection of water with the wavelength covering $0.8-1.7 \mu\mathrm{m}$.
The JWST Transiting Exoplanet ERS program \revise{\citep{Stevenson2016, Bean2018, 2023Natur.614..649J}} has measured the transmission spectra of WASP-39b using several instrument modes of JWST, including NIRCam \citep{Ahrer2023a}, NIRISS \citep{feinstein2023early}, NIRSpec G395H \citep{alderson2023early}, and NIRSpec PRISM \citep{Rustamkulov2023}.

The JWST NIRSpec PRISM spectra of WASP-39b collectively cover a wavelength range of $0.5 - 5.5 \mu\mathrm{m}$, which reveals the spectral features of \ce{H2O}, \ce{CO}, \ce{CO2}, \ce{SO2}, as well as that of \ce{Na} \citep{Rustamkulov2023}.
We use the data from \citet{Rustamkulov2023} observed by the NIRSpec PRISM instrument and reduced by \texttt{FIREFLy} pipeline.
We noted that the PRISM spectrum has relatively longer wavelength coverage, spanning more chemical species than others. This broader spectral range helps reduce degeneracy between molecular absorption features and enables a more comprehensive characterization of atmospheric composition and the exploration of limb-specific chemical, aerosols and thermal properties.

\begin{deluxetable*}{l c c c c c c c} [htb]
  \label{tab: retrieved para}
    \tablecaption{Retrieved results of JWST NIRSpec/PRISM observations using different hypotheses}
    \tablehead{& 1D Cloudy & 1D Patchy Cloudy & 2D Fixed modelA & 2D Fixed modelB & 2D Free modelA & 2D Free modelB}
    \startdata
    $R_{\text {p,1bar}}(R_{\text {J}})$ & $1.200_{-0.005}^{+0.005}$ & $1.195_{-0.006}^{+0.004}$ & $1.186_{-0.0039}^{+0.0042}$ & $1.186_{-0.0040}^{+0.0043}$ & $1.191_{-0.0039}^{+0.0037}$ & $1.189_{-0.0038}^{+0.0038}$  \\
    $\log Z(Z_\odot)$ & $2.140_{-0.060}^{+0.050}$ & $2.125_{-0.065}^{+0.055}$ & $1.948_{-0.072}^{+0.073}$ & $1.970_{-0.071}^{+0.073}$ & $1.846_{-0.083}^{+0.086}$ & $1.899_{-0.088}^{+0.084}$ \\
    $T(K)$ & $1280_{-50}^{+35}$ & $1260_{-45}^{+40}$ & - & - & - & - \\
    $T_{\text {morn }}$ & - & - & $1066_{-25}^{+39}$ & $1091_{-40}^{+47}$ & $874_{-97}^{+125}$ & $917_{-113}^{+126}$ \\
    $T_{\text {even }}$ & - & - & $1216^*$ & $1241^*$ & $1288_{-46}^{+36}$ & $1305_{-44}^{+33}$ \\
    $\log P_{\text {c}}$(bar) &  $-3.408_{-0.161}^{+0.193}$ & $-3.553_{-0.366}^{+0.158}$ & - & - & - & - \\
    $\log P_{\text {c,morn }}$ & - & - &  $-4.294_{-0.210}^{+0.151}$ & $-4.499_{-0.210}^{+0.237}$ & $-4.175_{-0.232}^{+0.264}$ & $-4.424_{-0.167}^{+0.252}$ \\
    $\log P_{\text {c,even }}$ & - & - &  $-0.762_{-1.766}^{+1.854}$ & $-2.546_{-0.251}^{+2.438}$ & $-2.731_{-0.104}^{+0.099}$ & $-2.737_{-0.105}^{+0.102}$ \\
    $\gamma$ &$-9.942_{-1.143}^{+0.985}$ & $-10.500_{-1.306}^{+1.138}$ & - & - & - & - \\
    $\gamma_{\text {morn }}$ & - & - & $-9.463_{-1.859}^{+2.151}$ & $-9.629_{-1.674}^{+1.938}$ & $-11.62_{-1.831}^{+1.786}$ & $-11.24_{-1.916}^{+1.831}$ \\
    $\gamma_{\text {even }}$ & - & - & $1.626_{-2.591}^{+0.271}$ & $-1.307_{-10.050}^{+3.072}$ & $-5.525_{-7.191}^{+5.301}$ & $-5.594_{-7.422}^{+5.264}$ \\
    $\log A$ & $1.589_{-0.265}^{+0.234}$ & $1.567_{-0.267}^{+0.229}$ & - & - & - & - \\
    $\log A_{\text {morn }}$ & - & - & $3.712_{-0.459}^{+0.445}$ & $3.529_{-0.504}^{+0.440}$ & $2.897_{-0.491}^{+0.475}$ & $3.105_{-0.443}^{+0.424}$ \\
    $\log A_{\text {even }}$ & - & - & $2.335_{-2.629}^{+0.221}$ & $-0.776_{-2.356}^{+3.154}$ & $-2.115_{-1.231}^{+2.034}$ & $-2.068_{-1.267}^{+1.897}$ \\
    $\mathrm{C} / \mathrm{O}$ & $0.677_{-0.041}^{+0.024}$ & $0.651_{-0.104}^{+0.035}$ & $0.456_{-0.091}^{+0.099}$ & - & $0.62_{-0.051}^{+0.032}$ & -  \\
    $\mathrm{C} / \mathrm{O}_{\text {morn }}$ & - & - & - & $0.365_{-0.127}^{+0.129}$ & - & $0.375_{-0.150}^{+0.148}$  \\
    $\mathrm{C} / \mathrm{O}_{\text {even }}$ & - & - & - & $0.606_{-0.146}^{+0.063}$ & - &  $0.649_{-0.040}^{+0.032}$  \\
    $\phi$ & - & $0.92_{-0.09}^{+0.06}$ & - & - & - & - \\
    \hline
    $\ln \mathcal{Z}$ & $1317.2 \pm 0.30$ & $1315.47 \pm 0.30$ & $1335.52 \pm 0.32$ & $1334.83 \pm 0.32$ & $1335.13 \pm 0.31$ & $1334.56 \pm 0.32$ \\
    $\chi^2/N$ & \revise{3.60} & \revise{3.60} & \revise{3.37} & \revise{3.37} & \revise{3.39} & \revise{3.38} \\
    \enddata
    \tablecomments{Free and Fixed models differ in their treatment of limb temperature differences: Free models treat the temperature difference as a free parameter, while Fixed models constrain it using the shallow water model assumption of a 150 K contrast. In Fixed models, the evening limb temperature (denoted by $^*$) is calculated by adding the retrieved morning limb temperature to the fixed temperature difference, rather than being an independently retrieved parameter as in Free models. Model A and Model B differ in their treatment of C/O: Model A assumes a homogeneous C/O ratio across both limbs, while Model B allows for a distinct C/O ratio in each limb.}
  \end{deluxetable*}

Table~\ref{tab: retrieved para} presents retrieved results from JWST NIRSpec/PRISM observations for an exoplanet's atmosphere under different hypotheses and models, which include 1D cloudy, 1D patchy cloud, Fixed model A/B, and Free model A/B. The parameters shown include physical, chemical, and atmospheric properties derived with statistical uncertainties.

The reference radius $R_{\text {p,1bar}}$ presents the radius of WASP-39 b's atmosphere. The retrieved values of $R_{\text {p,1bar}}$ across models range from 1.186 to $1.2\,R_\mathrm{Jup}$ with small uncertainties. Models are in good agreement with this parameter, indicating consistent radius retrieval regardless of the hypothesis.
$\log Z$ represents the atmospheric metallicity relative to solar values. 1D cloudy and patchy models exhibit slightly higher metallicity (2.1 dex) compared to 2D models (ranging from $\sim 1.86$ to $1.91$ dex), suggesting that differences in atmospheric assumptions influence this parameter.

The temperatures from the 1D model are 1280 K and 1260 K for the cloudy and patchy cloudy models, respectively, indicating a higher global temperature compared to the equilibrium temperature of 1116 K. The 2D fixed models show lower temperatures in both limbs. In contrast, the free models allow for limb-specific retrievals, revealing significant differences, with the morning limb being colder and the evening limb warmer than those in the corresponding fixed models.

The 1D models and 2D free model A show super-solar values for the global C/O ratio, whereas the 2D fixed model A results in a sub-solar value of $0.456^{+0.099}_{-0.091}$. The fixed and free model B can describe a diverse C/O ratio across the two limbs. In the fixed model B, the C/O ratio for the morning side is C/O$_{\text{morn}} = 0.365_{-0.127}^{+0.129}$, while for the evening side it is C/O$_{\text{even}} = 0.606_{-0.146}^{+0.063}$. Although these median values differ, there is a statistical possibility that they could be the same when considering the dispersion of C/O$_{\textit{i}}$ ($i = \text{morn, even}$), which is consistent with the sub-solar value obtained from fixed model A. However, the free model B predicts a more diverse C/O ratio across the limbs, with $0.375_{-0.150}^{+0.148}$ for morning and $0.649_{-0.040}^{+0.032}$ for evening.

The retrieved Bayesian evidence ($\ln \mathcal{Z}$) values for the various atmospheric models reveal a clear hierarchy in their ability to explain the JWST NIRSpec/PRISM observations. Fixed Model A achieves the highest evidence ($\ln \mathcal{Z} = 1335.52 \pm 0.32$) and serves as the benchmark for comparison. Models such as Fixed Model B ($\Delta \ln \mathcal{Z} = -0.69$), Free Model A ($\Delta \ln \mathcal{Z} = -0.39$), and Free Model B ($\Delta \ln \mathcal{Z} = -0.96$) exhibit negligible differences in Bayesian evidence ($\Delta \ln \mathcal{Z} < 1$), indicating that they are statistically indistinguishable from Fixed Model A in their explanatory power. In contrast, the simpler 1D Cloudy ($\Delta \ln \mathcal{Z} = -18.32$) and 1D Patchy Cloudy ($\Delta \ln \mathcal{Z} = -20.05$) models are strongly disfavored ($\Delta \ln \mathcal{Z} \geq 5$), suggesting that they fail to capture the necessary atmospheric complexity. \revise{The comparison of reduced chi-squared values demonstrates distinct advantages for 2D atmospheric models. While 1D models yield a significantly higher value of 3.60, their 2D counterparts show substantially improved fits ($3.37 \leq \chi^2/N \leq 3.39$)}\footnote{The relatively high $\chi^2$ values arise because part of the data suffer from detector saturation. Inflating the errors of these data points by a factor of 1000 reduces the reduced $\chi^2$ to 2.7.}. \revise{Among these 2D models, the chi-squared values are nearly identical, with no model showing statistical preference over others, suggesting} flexibility in model choice depending on the scientific objectives. This analysis highlights the importance of incorporating limb-specific atmospheric dynamics for accurate retrievals of exoplanetary atmospheres.

\begin{figure*}[htb]
  \centering
  \includegraphics[width=0.9\textwidth, angle=0]{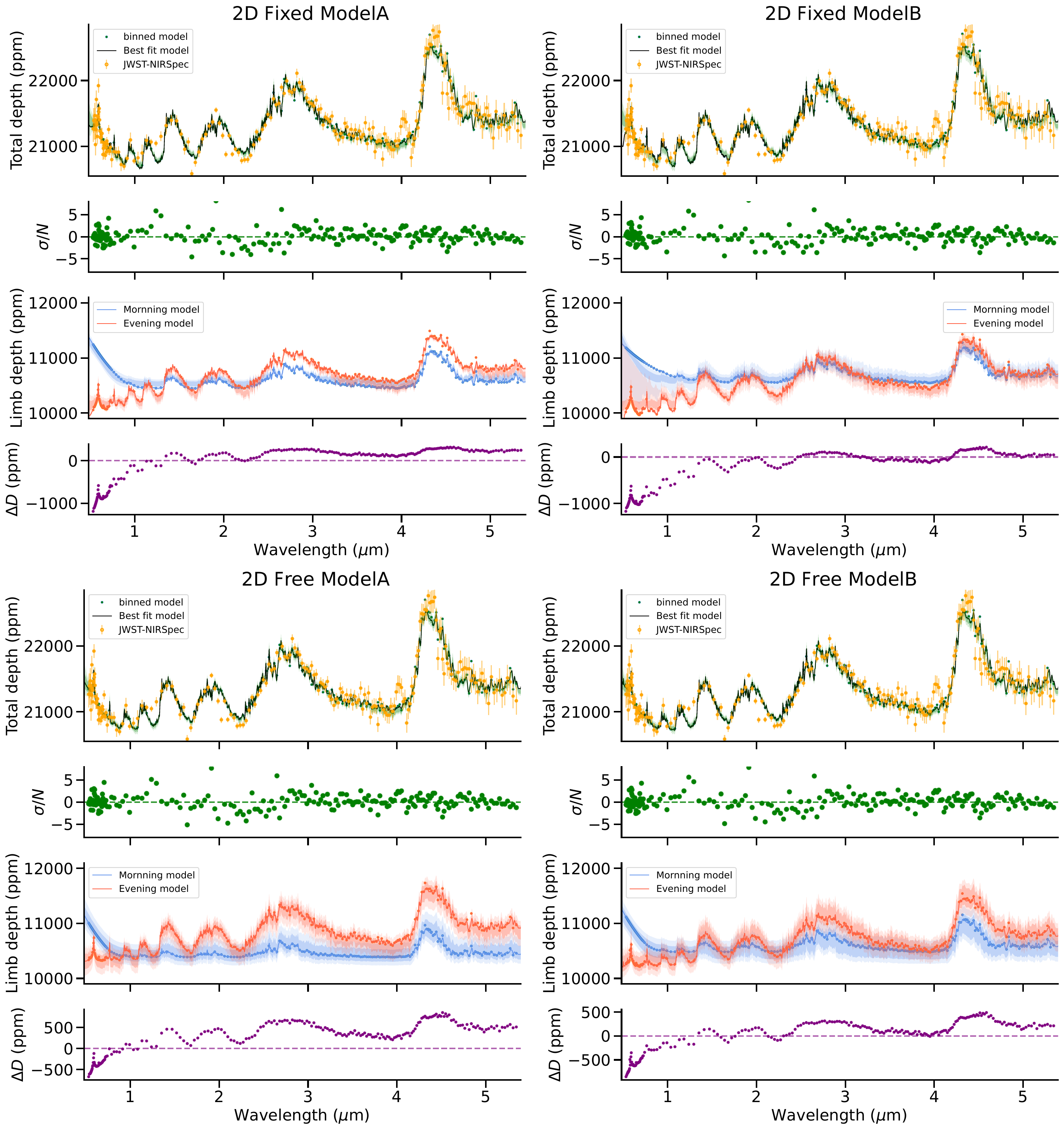}
  \caption{The retrieved transmission spectra for four 2D atmospheric models (2D Fixed Model A, 2D Fixed Model B, 2D Free Model A, and 2D Free Model B) derived from JWST NIRSpec/PRISM observations. Each panel represents the results for one of the models, divided into four subplots illustrating: (1) the total observed transit depth; (2) the normalized residuals; (3) the limb-specific (morning and evening) contributions to the transmission spectrum; and (4) the difference between morning and evening limb depths $\Delta D$.}
  \label{fig:transpec_jwst}
\end{figure*}

Fig~\ref{fig:transpec_jwst} displays retrieved transmission spectra for four 2D atmospheric models (2D Fixed Model A, 2D Fixed Model B, 2D Free Model A, and 2D Free Model B) derived from JWST NIRSpec/PRISM observations.

The top row of each panel shows the binned model points, best-fit model spectrum, and observed JWST-NIRSpec data points (yellow dots) with their observational error. Across all models, the fits align closely with the observed spectrum, reproducing prominent spectral features such as water absorption bands around $1.4 \, \mu\mathrm{m}$ and $2.7 \, \mu\mathrm{m}$, and \ce{CO2} near $4.3 \, \mu\mathrm{m}$. While the differences between models appear subtle in this plot, further insights can be gained from the residuals and limb-specific contributions.

The second row highlights the residuals normalized by the noise for each wavelength bin. For all models, the normalized residuals are generally distributed around zero, with most points falling within $\pm 2\sigma$. This indicates good statistical agreement between the models and the data. However, a few notable outliers, particularly between $0.7 - 1.9 ~\mu\mathrm{m}$, may be attributed to detector saturation effects \citep{Rustamkulov2023,Carter2024}.

The third row illustrates the individual contributions of the morning (blue) and evening (red) limbs to the total transmission spectrum. A consistent trend across all models reveals that the evening limb spectra exhibit stronger absorption features, indicative of a clearer atmosphere. In contrast, the morning limb spectra appear more muted and flattened, with a steep slope at shorter wavelengths, suggesting strong haze scattering.
Therefore, all four models predict significant differences in transmission depths between the morning and evening limbs in the visible band.
However, the models diverge at specific absorption bands, such as \ce{CO2} at $4.3~\mu\mathrm{m}$ and \ce{H2O} at $2.9 \, \mu\mathrm{m}$, where the free models predict larger differences compared to the fixed models.

The bottom row directly quantifies the limb difference ($\Delta D = D_{\text{even}} - D_{\text{morn}}$) as a function of wavelength. All models show significant variations in $\Delta D$, reaching up to $\sim 1000 \, \mathrm{ppm}$ in several regions, with pronounced differences near $2.9 \, \mu\mathrm{m}$ and $4.3 \, \mu\mathrm{m}$. These results highlight the critical role of limb-specific dynamics in accurately interpreting the transmission spectrum.

To facilitate comparison with the canonical method, we also apply our retrievals to the previously published dataset, which combines HST, VLT, and Spitzer observations presented in \citet{Wakeford2017} (hereafter referred to as the pre-JWST dataset). This dataset has been widely used to investigate both the properties of the WASP-39 system and atmospheric models \citep{kirk2019lrg, Fairman2024}.

As aforementioned, the full JWST NIRSpec/PRISM data is affected by saturation at $0.9-2.3\,\mu\mathrm{m}$. \citet{Rustamkulov2023} attempted to recover these data; however, \citet{Carter2024} and \citet{Espinoza2024} used only the portion of the spectrum with wavelengths longer than $2\,\mu\mathrm{m}$. Here we also use this truncated spectrum for our retrievals, which we refer to as the JWST cutoff.

\section{Comparison among Datasets and Models}\label{sec:discuss}

Here we further compare our findings with previous retrieval analyses of WASP-39b, investigate the impact of accounting for differences in temperature, composition, and cloud-haze properties between the morning and evening limbs, and discuss the limitations of our retrieval models.

\subsection{Temperature}
\begin{figure*}[htb]
  \centering
  \includegraphics[width=\textwidth, angle=0]{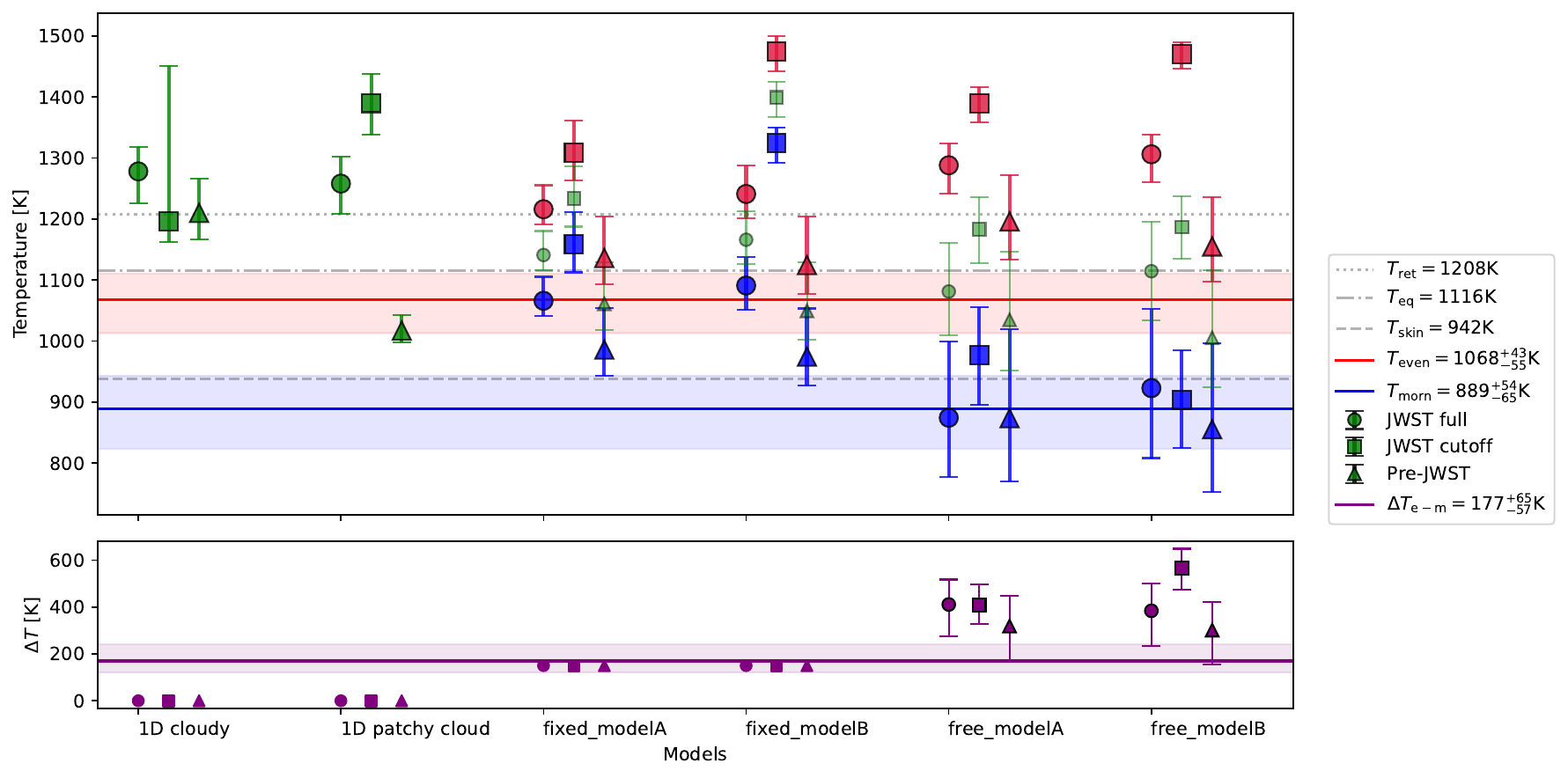}
  \caption{
  Comparison retrieved temperature of WASP-39 b of six hypotheses, arranged from left to right: 1D homogeneous clouds, 1D patchy clouds; 2D fixed $150\,\mathrm{K}$ temperature difference with homogeneous C/O and inhomogeneous cloud haze, both C/O and aerosol are inhomogeneous; 2D free temperature difference with homogeneous C/O and inhomogeneous cloud haze, both C/O and aerosol are inhomogeneous.
  Each hypothesis is marked using different symbols corresponding to the datasets used: circles represent the JWST/PRISM full dataset, while JWST cutoff dataset and the pre-JWST dataset are shown in squares and triangles respectively.
  The upper panel describes the retrieved temperature parameter, a global temperature(in green) for 1D hypotheses, while morning temperature(in blue), evening temperature(in red) and the averaged temperature(in green) for 2D hypotheses. And the bottom shows the temperature difference correspondingly.
  Three grey dotted, dashdot and dashed line present the reference retrieved temperature, equilibrium temperature and skin temperature from   \citet{Welbanks2022} respectively. The red, blue and purple solid line with shaded areas present the retrieved median and $1\sigma$ confidence intervals result from  \citet{Espinoza2024}.
  }
  \label{fig:temp_distribution}

\end{figure*}

Temperature differences are widely regarded as the primary factor driving variations between the morning and evening transmission spectra in studies of limb asymmetry \citep{Powell2019,Caldas2019,Ehrenreich2020}. However, evidence of limb asymmetry in low-resolution transmission spectroscopy of hot Jupiters has been observed simply for WASP-39b and WASP-107b. Until \citet{Espinoza2024} extracted individual spectra for each limb, no direct measurements of morning and evening limb temperatures had been reported for WASP-39b.

To address this, we analyze the temperatures obtained from our retrievals, focusing on both the global temperature and the temperature differences between the morning and evening limbs.

On the global scale, many previous works and 1D retrievals are available for comparison, as summarized in Figure~\ref{fig:temp_distribution}. The top panel presents the absolute values of global temperatures retrieved under different hypotheses, shown as green dots with error bars. For 2D models, we calculate the average of the morning and evening limb temperatures to represent the global atmospheric state. These values are compared against three theoretical benchmarks from \citet{Welbanks2022}, represented by horizontal grey lines: the equilibrium temperature \revise{(dotted, $T_\text{eq}=\left[f\left(1-A_B\right)\left(\frac{R_\text{star}}{2a}\right)^2\right]^{\frac{1}{4}}$) assuming no bond albedo ($A_B = 0$) and full energy redistribution;} the skin temperature (dash-dot) \revise{given by $T_\text{skin} = 2^{-1/4} T_\text{eq}$ which represents the temperature at the outer atmospheric layer where the optical depth is low, making it transparent to incoming stellar radiation while being heated primarily by the planet's outgoing thermal radiation in a gray atmosphere model;} and the retrieved limb temperature (dashed) \revise{at pressure of 100 mbar from \citet{welbanks2019degeneracies}}.

In 1D tests, we find that the JWST full dataset is less sensitive to variations in cloud coverage, while the pre-JWST dataset retrieves significantly lower temperatures when considering a patchy cloud model.
This indicates that the treatment of cloud coverage significantly affects temperature in 1D retrievals which imply inhomogeneous clouds treatment.

For 2D cases, the retrieved global temperatures remained consistently close to WASP-39 b's equilibrium temperature ($1116\,\mathrm{K}$) across the datasets, except for the JWST cutoff dataset. The temperatures retrieved from pre-JWST dataset were consistently yields lower temperatures than those from the other two datasets.
This discrepancy might be related to the different wavelength coverage between pre-JWST and JWST observations, though the specific mechanism requires further investigation. Different wavelength ranges could probe different atmospheric layers or be influenced by different opacity sources, potentially affecting the temperature retrievals.

Among the tested hypotheses, Fixed Model A demonstrates the most consistent temperature retrievals across all datasets. This suggests that Fixed Model A effectively balances the representation of cloud and temperature variations while maintaining stable global temperature estimates.

In our model, the temperatures of both the morning and evening limbs are retrieved, providing valuable insights into the physical and chemical processes occurring in the atmospheres of distant exoplanets. Both the absolute temperatures and the temperature differences between these limbs are important for understanding atmospheric dynamics and energy distribution.

To date, only one study by  \citet{Espinoza2024} reported limb-specific retrieved temperatures for WASP-39 b. Fig~\ref{fig:temp_distribution} illustrates their results, showing the median temperature and $1 \sigma $ uncertainty for both the morning ($T_\text{morn}=889^{+54}_{-65}\,\mathrm{K}$ ) and evening limbs ($T_\text{even}=1068^{+43}_{-55}\,\mathrm{K}$), as well as the temperature difference ($\Delta T_\text{e-m}=177^{+65}_{-57}\,\mathrm{K}$). These values are represented with solid lines and shaded regions in blue, red, and purple, respectively. This work offers a benchmark for limb-specific temperature retrievals.

The JWST cutoff dataset and the reference data from \citet{Espinoza2024} are identical, both using the $>2 \, \mu\mathrm{m}$ part of the JWST NIRSpec/PRISM observations. However, the retrieved temperatures from the JWST cutoff dataset exhibit the largest deviation from the reference values. In contrast, the pre-JWST results show slightly closer agreement with the observations compared to those obtained from the full JWST dataset.

Among the four 2D models, Fixed Model A demonstrates greater robustness across different datasets, though it consistently overestimates temperatures by approximately 150 K relative to the observations. On the other hand, Fixed/Free Model B exhibits higher sensitivity to the choice of dataset, likely due to the inclusion of one more C/O ratio as a free parameter. This added complexity increases the model's reliance on the quality of the data.

\revise{Our shallow-water model predicts a temperature difference of approximately $150~\mathrm{K}$ between morning and evening terminators. While this difference aligns with the findings of \citet{Espinoza2024}, we note that our absolute temperatures differ from their results. Free models that allow independent variation of morning and evening limb temperatures recover larger temperature differences (approximately $388~\mathrm{K}$ and $414~\mathrm{K}$) and provide better fits to specific spectral features like CO2. While allowing both limb temperatures to vary freely can introduce additional parameter degeneracies, the improved spectral fits suggest these larger temperature contrasts may reflect actual atmospheric conditions. These findings highlight the importance of considering both constrained and unconstrained approaches when interpreting atmospheric retrievals.}

\subsection{Metallicity and C/O ratio }

The atmospheric composition of WASP-39b is characterized by two primary parameters: metallicity ($\log Z$) and the C/O ratio.
In all models, metallicity is treated as uniform across the atmosphere, while the C/O ratio varies between homogeneous (in Model A) and inhomogeneous (in Model B) assumptions.
\begin{figure}[htb]
  \centering
  \includegraphics[width=\columnwidth, angle=0]{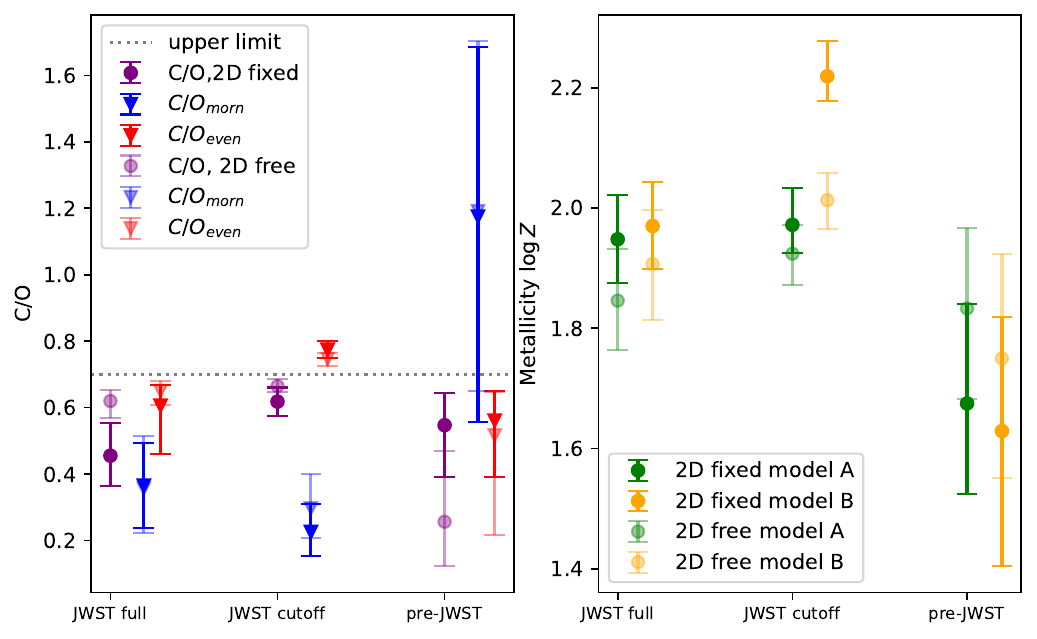}
  \caption{The retrieved results of composition parameters of three datasets and four models. \revise{The C/O upper limit of 0.7 is reported by \citep{Rustamkulov2023} through the full JWST NIRSpec/PRISM observation.}}
  \label{fig:composition}
\end{figure}

For the JWST full dataset, both Fixed Model A and B suggest that the C/O ratios of WASP-39 b's morning and evening limbs are consistent with each other. Fixed Model A yields a C/O ratio of $0.46_{-0.09}^{+0.10}$ and Fixed Model B provides limb-specific value of C/O$_\text{morn}=0.37 ^{+0.13}_{-0.33}$ and C/O$_\text{even}=0.60 ^{+0.06}_{-0.15}$. These results align with the previous findings, including \citet{Rustamkulov2023}, who reported C/O$ \approx 0.3-0.5$ and \citep{Espinoza2024}, who found C/O$_\text{morn}=0.57 ^{+0.17}_{-0.23}$, C/O$_\text{even}=0.58 ^{+0.13}_{-0.16}$.
In contrast, the Free Models retrieved more divergent results. Free Model A suggest a super-solar C/O ratio of $0.62^{+0.03}_{-0.04}$ while Free Model B indicates a higher C/O in the evening limb (C/O$_\text{even}=0.65 ^{+0.03}_{-0.04}$) compared to the morning limb (C/O$_\text{morn}=0.36 ^{+0.16}_{-0.13}$).

The discrepancies become more pronounced when analyzing the JWST cutoff dataset using Model B. Both fixed and free versions of Model B result in C/O$_\text{even} \approx 0.77$ exceeding the upper limit of 0.7 reported in \citet{Rustamkulov2023}.

For the pre-JWST dataset, Fixed Model B retrieves a very high median $ \text{C/O}_\text{even} $ ratio; however, the error ranges for both Fixed Model A and B overlap around the solar value, supporting the possibility of a homogeneous atmospheric composition. In contrast, the Free Models do not favor this possibility. Moveover, Free Model A yields a relatively low C/O ratio of $ 0.26^{+0.21}_{-0.12} $, closely aligning with the value of $ 0.31^{+0.08}_{-0.05} $ retrieved by the ATMO model \citep{Wakeford2017}.

These results show the variability introduced by model selection and dataset. Across all cases, the data favor a homogeneous atmospheric composition for WASP-39b's morning and evening limbs. While theoretically capable of describing inhomogeneous composition, Model B introduces interpretative challenges due to its reliance on two free C/O parameters. This can lead to degenerate or contradictory results, as evidenced by the retrievals from the JWST cutoff and pre-JWST datasets. We argue that the use of Model A, which assumes homogeneity, is more robust and reliable, especially when working with datasets of limited quality.

With respect to atmospheric composition, a further limitation of the work on WASP-39 b in this paper is the inadequate description of \ce{SO2}. Excess \ce{SO2} absorption was detected on WASP-39 b, which provided the first evidence of photochemically induced chemical disequilibrium on an exoplanet \revise{\citep{Tsai2023c,2024Natur.626..979P}}. We used only the chemical equilibrium model to retrieve the observations and did not take into account the additional injection of \ce{SO2} or non-chemical equilibrium, as that would introduce more parameters.

\subsection{Aerosol Properties}

We find that the distribution of clouds and haze on the morning and evening limbs of WASP-39 b exhibits significant asymmetry. Previous GCM studies have often suggested that such inhomogeneous distributions are primarily expected for hot gas-giants with equilibrium temperatures exceeding $1600\,\mathrm{K}$ \citep{Powell2024a}. However, our results demonstrate that even at lower temperatures, WASP-39 b shows remarkable differences in its cloud and haze properties between the two limbs. These differences have important implications for the planet's transmission spectrum: the deeper cloud distribution on the evening limb corresponds to clearer atmospheric conditions, allowing for enhanced absorption features in the spectrum, while the stronger scattering on the morning limb produces a steeper Rayleigh-like scattering slope, indicating the presence of smaller haze particles or higher-altitude aerosols.

\begin{figure*}[htb]
  \centering
  \includegraphics[width=0.8\textwidth, angle=0]{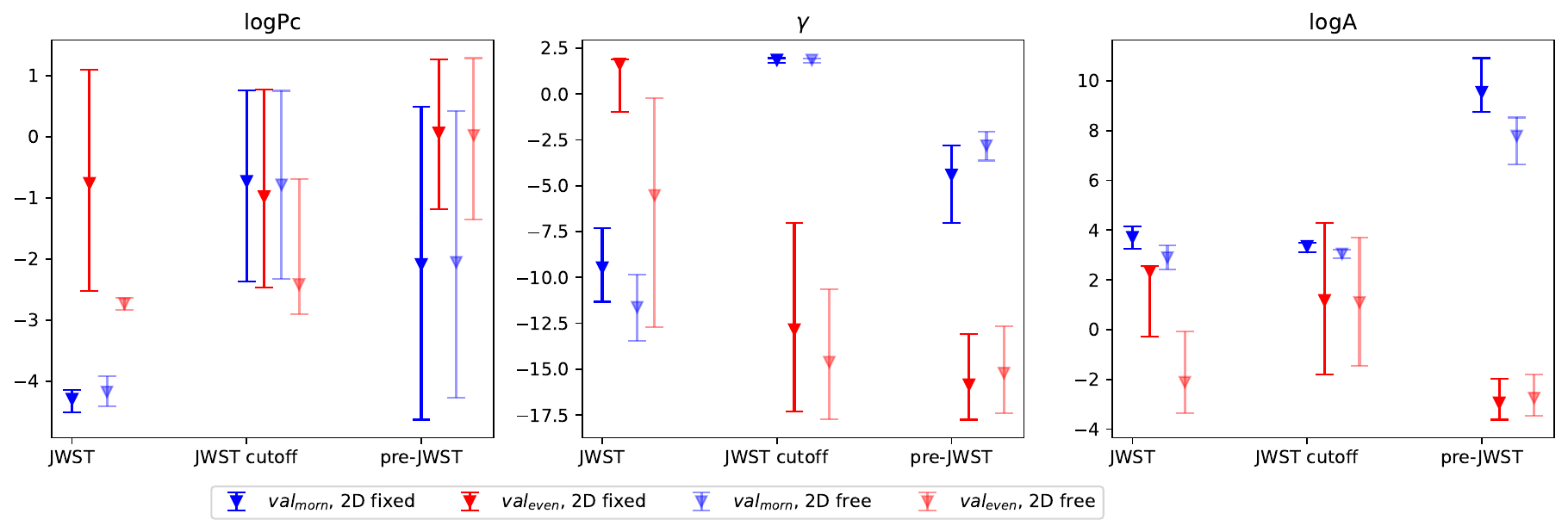}
  \caption{Comparison of retrieved results: cloud top pressure ($\log P_\mathrm{c}$\revise{[bar]}), Rayleigh-like scattering slope ($\gamma$), and logarithmic scattering enhancement factor ($\log A$) across JWST full, JWST cutoff, and Pre-JWST datasets. Results are shown for $\text{val}_\text{morn}$ (blue) and $\text{val}_\text{even}$ (red), under fixed (\revise{dark}) and free (light) models. Error bars represent uncertainties for each parameter.}
  \label{fig:aerosol}
\end{figure*}

The differences between datasets are highlighted by three parameters presented in Figure \ref{fig:aerosol}: cloud top pressure ($\log P_c$), scattering slope ($\gamma$), and the logarithmic scattering enhancement factor ($\log A$).

In the left panel ($\log P_c$), the JWST full dataset provides tighter constraints on cloud top pressures due to its broader and more continuous wavelength coverage. This dataset reveals that clouds are located deeper in the atmosphere on the evening limb, supporting the hypothesis of clearer atmospheric conditions \revise{than the morning limb}. In contrast, the JWST cutoff dataset, which excludes wavelengths below $2\, \mu\mathrm{m}$, does not distinguish the height difference between clouds on the morning and evening limbs effectively. \revise{This strongly suggests that the $0.6 - 2\, \mu\mathrm{m}$ range contributes critically to resolving the asymmetry. While in pre-JWST dataset which has relatively dense sampling in the $0.3 - 1.6\, \mu\mathrm{m}$, shows hints of the asymmetry despite large uncertainties. This suggests the atmospheric structure difference was potentially encoded in these wavelengths in early observations, though insufficient for robust characterization without additional longer wavelength coverage.}

\revise{Generally, additional optical wavelength coverage ($0.3 - 1.1\, \mu\mathrm{m}$) does not significantly improve constraints on cloud-top pressure ($\log P_c$ ) for most exoplanets\citep{Fairman2024}. This is because many exoplanet spectra only yield upper limits corresponding to non-detection of optically thick cloud decks. However, \citet{Fairman2024} specifically identify WASP-39b as a notable exception to this trend, demonstrating that optical wavelengths can substantially enhance cloud deck constraints when high-altitude, gray clouds are present.  We extends their findings in a crucial way by identifying that these high-altitude gray clouds show a specific morning-evening asymmetry - with high-altitude clouds predominantly present on the morning terminator while largely absent on the evening side. Our analysis not only confirms this exceptional case but extends their findings in a crucial way by revealing that these high-altitude gray clouds exhibit a pronounced morning-evening asymmetry—predominantly present on the morning terminator while largely absent on the evening side. This asymmetric distribution explains the enhanced diagnostic power of optical and short-wavelength near-IR observations in our dataset, as these wavelengths are particularly sensitive to the altitude contrast between the cloud-dominated morning and the relatively cloud-free evening terminators.}

In the middle panel ($\gamma$), the JWST full dataset suggests that haze particles on the morning limb produce a steeper Rayleigh-like scattering slope compared to the evening limb. However, this conclusion contrasts with the findings of the JWST cutoff and pre-JWST datasets, which indicate the opposite trend. These discrepancies are likely caused by parameter degeneracies and the under-constrained nature of the models in the latter datasets. We favor the conclusions drawn from the JWST full dataset due to its inclusion of critical data near $1\, \mu\mathrm{m}$, which is absent in the JWST cutoff dataset, and the higher density and continuity of its data compared to the sparse pre-JWST measurements.

In the right panel ($\log A$), all three datasets display a consistent trend, indicating stronger scattering on the morning limb and weaker haze scattering on the evening limb. This agreement reinforces the conclusion that the morning limb is characterized by a more optically active haze layer.

Despite the simplified parameterizations used in our model, these results are in good agreement with the asymmetry laws proposed by \cite{Powell2024a}. The observed differences in cloud and haze properties between the morning and evening limbs support the hypothesis that atmospheric temperature gradients and stellar radiation can drive asymmetric distributions of clouds and haze, even for exoplanets with equilibrium temperatures below $1600\,\mathrm{K}$.

\section{Conclusion}\label{sec:conclude}

This study investigates the temperature, composition, and cloud-haze asymmetries between the morning and evening atmospheres of the hot Saturn WASP-39 b. We introduce a new fixed temperature difference 2D retrieved model and evaluated the necessity, applicability, and robustness of four 2D models (2D Fixed Model A/B and 2D Free Model A/B) retrievals under simplified atmospheric assumptions:
\begin{itemize}
  \item Fixed Model A: Implements a pre-calculated morning-evening temperature difference of 150 K from shallow water model calculations, with a homogeneous C/O ratio across both limbs
  \item Fixed Model B: Maintains the fixed 150 K temperature difference while allowing distinct C/O ratios for each limb
  \item Free Model A: Treats the morning-evening temperature difference as a free parameter, with a homogeneous C/O ratio
  \item Free Model B: Combines free temperature difference with distinct C/O ratios for each limb
\end{itemize}

Our key findings demonstrate that the shallow water model effectively predicts a morning-to-evening temperature difference of $150\,\mathrm{K}$, consistent with observations \citep{Espinoza2024}.The results from the 2D models show that\revise{ under the assumption of a uniform metallicity across the limbs, there is no statistical evidence for inhomogeneity in the C/O ratio between the limbs}. However, the morning limb features higher-altitude clouds and stronger haze scattering, while the evening limb is relatively clear, corresponding to deeper clouds.

\revise{Bayesian evidence strongly supports the necessity of 2D models when interpreting JWST transmission spectra, confirming that 1D models fail to capture the atmospheric asymmetries. Among the four 2D models tested, Bayesian evidence could not distinguish their performance based on the total transmission spectrum, indicating that both fixed and free approaches remain statistically viable. The 2D Free Models recover larger morning-to-evening temperature differences than our SWAMPE model predicts, while offering improved fits to certain spectral features. The 2D Fixed Model A, constrained by the $150\,\mathrm{K}$ temperature contrast from our shallow-water model, provides consistent results across different datasets. However, we note that both approaches yield similar results for most atmospheric parameters as shown in Figures 4 and 5, suggesting that the choice between fixed and free models represents a trade-off between prior constraints and data-driven flexibility.}

The findings of this study demonstrate that atmospheric asymmetries, such as those observed on WASP-39 b, are not limited to ultra-hot Jupiters with equilibrium temperatures above $1600\,\mathrm{K}$. Significant morning-evening variations in temperature, clouds, and haze can also occur on moderately warm planets, broadening our understanding of atmospheric dynamics across a wider range of tidally locked planets.

For exoplanets where independent morning and evening limb spectra are not currently accessible due to observational constraints, models such as the shallow water model used here or more advanced 3D GCMs can predict limb temperature contrasts.
By combining these predictions with high-precision transmission spectra, such as those obtained from JWST, we can effectively probe atmospheric asymmetries indirectly.

Extending this approach to a larger sample of exoplanets will allow for systematic exploration of atmospheric inhomogeneities across diverse planetary types and conditions. Such work can shed light on the physical processes driving cloud and haze distributions and enhance our understanding of exoplanet climate dynamics.

While this work advances our understanding of atmospheric asymmetries on WASP-39 b, the models remain simplified. We assumed isothermal temperature profiles, chemical equilibrium, and parameterized cloud and haze properties. \revise{The choice of temperature profile parameterization, particularly the isothermal assumption, can introduce systematic biases in atmospheric retrievals \citep{Welbanks2022}, which represents a key limitation of our approach. }In the Fixed Models, the temperature difference of $150\,\mathrm{K}$ was applied, without accounting for potential variations in limb temperature based on retrieval results.

Future studies should aim to address the fixed temperature \revise{difference used in Fixed Model by developing a dynamic function to allow for variable limb temperature differences, rather than allowing them to vary independently as in Free Model. This could involve creating relationships between equilibrium temperature and limb temperature differences derived from GCM studies.} Additionally, independent observational evidence for limb spectral differences, particularly at visible wavelengths and near strong absorption bands, will be crucial for further distinguishing between competing 2D models and confirming the correct interpretation of atmospheric asymmetries.

We note that a method for extracting morning and evening independent spectra in light-variation curves was recently proposed \citep{Espinoza2021} and applied to WASP-107 b \citep{Murphy2024} and WASP-39 b \citep{Espinoza2024}. This method begins to account for the effect of limb inhomogeneities at an early stage, starting with the processing of light curves into spectra; however, this approach requires highly precise transit timing, meaning that the planet's orbital parameters must be well-constrained. In the future, for larger samples of planetary targets to be studied, further determination of three-dimensional orbits of the planets through the astrometry mission  \citep{Ji2022,2024ChJSS..44..193J} is pending in addition to the current inception methods to provide more information for such work. Until that time, our work can help address the challenge of more plausibly interpreting the synthetic spectra of transiting planets within the framework of the limb asymmetry model.

\section*{Acknowledgements}
We thank the anonymous reviewer for their insightful comments that improved the quality of this work. The JWST data presented in this article were obtained from the Mikulski Archive for Space Telescopes (MAST) at the Space Telescope Science Institute. The specific observations analyzed can be accessed via \dataset[doi:10.17909/yqe4-fe42]{https://doi.org/10.17909/yqe4-fe42}. This work is financially supported by the National Natural Science Foundation of China (grant Nos. 12033010 and 11773081), the Strategic Priority Research Program on Space Science of the Chinese Academy of Sciences (grant No. XDA 15020800), and the Foundation of Minor Planets of the Purple Mountain Observatory.


\vspace{5mm}

\software{astropy  \citep{TheAstropyCollaboration2013,TheAstropyCollaboration2018},
           \swampe  \citep{Landgren2023},
          \texttt{PLATON}  \citep{Zhang2019,Zhang2020a}
          }

\appendix
\section{additional tables and figures}
\begin{figure}[htb]
  \centering
  \includegraphics[width=0.7\textwidth, angle=0]{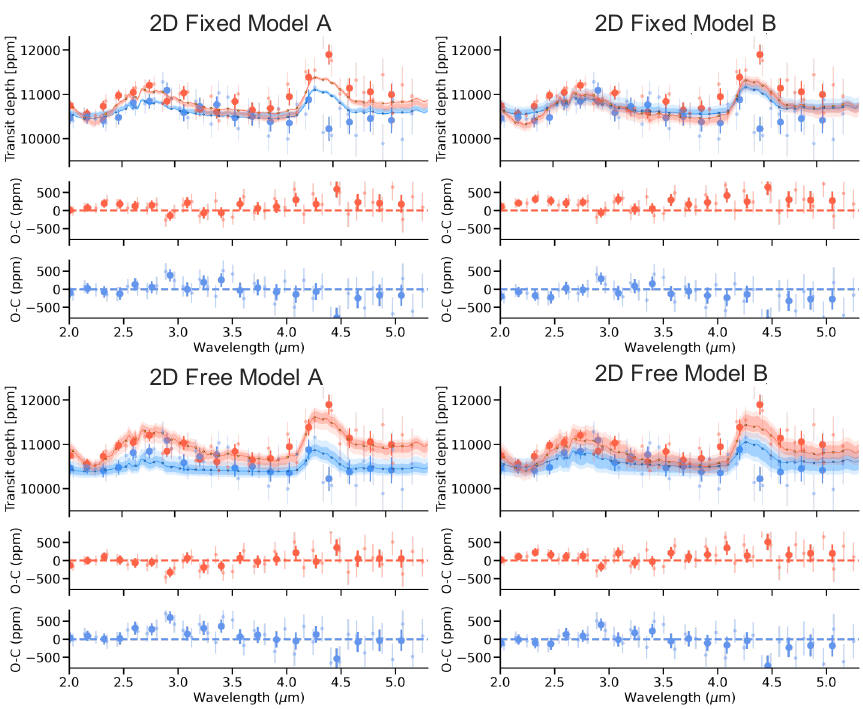}
  \caption{Supplement to Fig.~\ref{fig:transpec_jwst}. Retrieved limb-specific spectra for four 2D atmospheric models derived from JWST NIRSpec/PRISM observations, with wavelengths less than $2\, \mu\mathrm{m}$ removed. This allows a direct comparison with the observational limb spectra reported by \citet{Espinoza2024}.}
  \label{fig:obs_spec}
\end{figure}

In this Appendix, we present the more detailed retrieved limb-specific spectra for four 2D atmospheric models (Fig.~\ref{fig:obs_spec}). The best-fit spectra are derived from the full JWST NIRSpec/PRISM observations, with wavelengths shorter than 2 $\mu\mathrm{m}$ removed when plotting to facilitate a direct comparison with the observational limb spectra reported by \citet{Espinoza2024}, which uses cutoff data from JWST NIRSpec/PRISM observations.
In Fig.~\ref{fig:obs_spec}, the red curves represent the evening-side spectra, while the blue curves represent the morning-side spectra. Solid lines with shaded regions show the model retrieval results, with the shaded areas indicating uncertainties. Observational data points, displayed with error bars, correspond to the limb spectra obtained at high (semi-transparent points) and low (opaque points) spectral resolutions \citep{Espinoza2024}.
The four 2D models generally agree with the observational data across most wavelengths, except at the \ce{CO2} absorption feature around 4.3 $\mu\mathrm{m}$. Observations reveal a limb difference of up to 2000 ppm at this wavelength. The condensate cloud models, photochemical haze treatments, and equilibrium chemistry with transport-induced disequilibrium used by \citet{Espinoza2024} fail to fit this feature. 
\revise{In our 2D Free model, the retrieval indicates a significantly higher evening-limb temperature to match this strong signal. Rather than necessarily representing an overestimation, this larger temperature difference may be capturing actual atmospheric properties not fully accounted for in our shallow-water model, particularly given the improved fit to CO2 features shown in Fig.~\ref{fig:obs_spec}.}

\bibliography{ms}

\begin{thebibliography}{}
\expandafter\ifx\csname natexlab\endcsname\relax\def\natexlab#1{#1}\fi
\providecommand{\url}[1]{\href{#1}{#1}}
\providecommand{\dodoi}[1]{doi:~\href{http://doi.org/#1}{\nolinkurl{#1}}}
\providecommand{\doeprint}[1]{\href{http://ascl.net/#1}{\nolinkurl{http://ascl.net/#1}}}
\providecommand{\doarXiv}[1]{\href{https://arxiv.org/abs/#1}{\nolinkurl{https://arxiv.org/abs/#1}}}

\bibitem[{Ahrer {et~al.}(2023)Ahrer, Stevenson, Mansfield, Moran, Brande,
  Morello, Murray, Nikolov, {Petit dit de la Roche}, Schlawin, Wheatley, Zieba,
  Batalha, Damiano, Goyal, Lendl, Lothringer, Mukherjee, Ohno, Batalha,
  Battley, Bean, Beatty, Benneke, {Berta-Thompson}, Carter, Cubillos, Daylan,
  Espinoza, Gao, Gibson, Gill, Harrington, Hu, Kreidberg, Lewis, Line,
  {L{\'o}pez-Morales}, Parmentier, Powell, Sing, Tsai, Wakeford, Welbanks,
  Alam, Alderson, Allen, Anderson, Barstow, Bayliss, Bell, Blecic, Bryant,
  Burleigh, Carone, Casewell, Changeat, Chubb, Crossfield, Crouzet, Decin,
  D{\'e}sert, Feinstein, Flagg, Fortney, Gizis, Heng, Iro, Kempton, Kendrew,
  Kirk, Knutson, Komacek, Lagage, Leconte, {Lustig-Yaeger}, MacDonald, Mancini,
  May, Mayne, Miguel, {Mikal-Evans}, Molaverdikhani, Palle, Piaulet, Rackham,
  Redfield, Rogers, Roy, Rustamkulov, Shkolnik, Sotzen, Taylor, Tremblin,
  Tucker, Turner, {de Val-Borro}, Venot, \& Zhang}]{Ahrer2023a}
Ahrer, E.-M., Stevenson, K.~B., Mansfield, M., {et~al.} 2023, Nature, 614, 653,
  \dodoi{10.1038/s41586-022-05590-4}

\bibitem[{Alderson {et~al.}(2023)Alderson, Wakeford, Alam, Batalha, Lothringer,
  Adams~Redai, Barat, Brande, Damiano, Daylan, {et~al.}}]{alderson2023early}
Alderson, L., Wakeford, H.~R., Alam, M.~K., {et~al.} 2023, Nature, 614, 664

\bibitem[{Allard {et~al.}(2016)Allard, Spiegelman, \& Kielkopf}]{Allard2016}
Allard, N.~F., Spiegelman, F., \& Kielkopf, J.~F. 2016, Astronomy \&
  Astrophysics, 589, A21, \dodoi{10.1051/0004-6361/201628270}

\bibitem[{Allard {et~al.}(2019)Allard, Spiegelman, Leininger, \&
  Molliere}]{Allard2019}
Allard, N.~F., Spiegelman, F., Leininger, T., \& Molliere, P. 2019, Astronomy
  \& Astrophysics, 628, A120, \dodoi{10.1051/0004-6361/201935593}

\bibitem[{Amundsen {et~al.}(2016)Amundsen, Mayne, Baraffe, Manners, Tremblin,
  Drummond, Smith, Acreman, \& Homeier}]{Amundsen2016}
Amundsen, D.~S., Mayne, N.~J., Baraffe, I., {et~al.} 2016, Astronomy and
  Astrophysics, \dodoi{10.1051/0004-6361/201629183}

\bibitem[{Bean {et~al.}(2018)Bean, Stevenson, Batalha, {Berta-Thompson},
  Kreidberg, Crouzet, Benneke, Line, Sing, Wakeford, Knutson, Kempton,
  D{\'e}sert, Crossfield, Batalha, de~Wit, Parmentier, Harrington, Moses,
  {Lopez-Morales}, Alam, Blecic, Bruno, Carter, Chapman, Decin, Dragomir,
  Evans, Fortney, Fraine, Gao, Mu{\~n}oz, Gibson, Goyal, Heng, Hu, Kendrew,
  Kilpatrick, Krick, Lagage, Lendl, Louden, Madhusudhan, Mandell, Mansfield,
  May, Morello, Morley, Nikolov, Redfield, Roberts, Schlawin, Spake, Todorov,
  Tsiaras, Venot, Waalkes, Wheatley, Zellem, Angerhausen, Barrado, Carone,
  Casewell, Cubillos, Damiano, de~{Val-Borro}, Drummond, Edwards, Endl,
  Espinoza, France, Gizis, Greene, Henning, Hong, Ingalls, Iro, Irwin, Kataria,
  Lahuis, Leconte, {Lillo-Box}, Lines, Lothringer, Mancini, Marchis, Mayne,
  Palle, Rauscher, Roudier, Shkolnik, Southworth, Swain, Taylor, Teske,
  Tinetti, Tremblin, Tucker, van Boekel, Waldmann, Weaver, \&
  Zingales}]{Bean2018}
Bean, J.~L., Stevenson, K.~B., Batalha, N.~M., {et~al.} 2018, Publications of
  the Astronomical Society of the Pacific, 130, 114402,
  \dodoi{10.1088/1538-3873/aadbf3}

\bibitem[{{Bell} {et~al.}(2024){Bell}, {Crouzet}, {Cubillos}, {Kreidberg},
  {Piette}, {Roman}, {Barstow}, {Blecic}, {Carone}, {Coulombe}, {Ducrot},
  {Hammond}, {Mendon{\c{c}}a}, {Moses}, {Parmentier}, {Stevenson},
  {Teinturier}, {Zhang}, {Batalha}, {Bean}, {Benneke}, {Charnay}, {Chubb},
  {Demory}, {Gao}, {Lee}, {L{\'o}pez-Morales}, {Morello}, {Rauscher}, {Sing},
  {Tan}, {Venot}, {Wakeford}, {Aggarwal}, {Ahrer}, {Alam}, {Baeyens},
  {Barrado}, {Caceres}, {Carter}, {Casewell}, {Challener}, {Crossfield},
  {Decin}, {D{\'e}sert}, {Dobbs-Dixon}, {Dyrek}, {Espinoza}, {Feinstein},
  {Gibson}, {Harrington}, {Helling}, {Hu}, {Iro}, {Kempton}, {Kendrew},
  {Komacek}, {Krick}, {Lagage}, {Leconte}, {Lendl}, {Lewis}, {Lothringer},
  {Malsky}, {Mancini}, {Mansfield}, {Mayne}, {Evans-Soma}, {Molaverdikhani},
  {Nikolov}, {Nixon}, {Palle}, {Petit dit de la Roche}, {Piaulet}, {Powell},
  {Rackham}, {Schneider}, {Steinrueck}, {Taylor}, {Welbanks}, {Yurchenko},
  {Zhang}, \& {Zieba}}]{2024NatAs...8..879B}
{Bell}, T.~J., {Crouzet}, N., {Cubillos}, P.~E., {et~al.} 2024, Nature
  Astronomy, 8, 879, \dodoi{10.1038/s41550-024-02230-x}

\bibitem[{Buchner {et~al.}(2014)Buchner, Georgakakis, Nandra, Hsu, Rangel,
  Brightman, Merloni, Salvato, Donley, \& Kocevski}]{Buchner2014}
Buchner, J., Georgakakis, A., Nandra, K., {et~al.} 2014, Astronomy \&
  Astrophysics, 564, A125, \dodoi{10.1051/0004-6361/201322971}

\bibitem[{Caldas {et~al.}(2019)Caldas, Leconte, Selsis, Waldmann, Bord{\'e},
  Rocchetto, \& Charnay}]{Caldas2019}
Caldas, A., Leconte, J., Selsis, F., {et~al.} 2019, Astronomy \& Astrophysics,
  623, A161, \dodoi{10.1051/0004-6361/201834384}

\bibitem[{Carter {et~al.}(2024)Carter, May, Espinoza, Welbanks, Ahrer,
  Alderson, Brahm, Feinstein, Grant, Line, Morello, O'Steen, Radica,
  Rustamkulov, Stevenson, Turner, Alam, Anderson, Batalha, Battley, Bayliss,
  Bean, Benneke, {Berta-Thompson}, Brande, Bryant, Burleigh, Coulombe,
  Crossfield, Damiano, D{\'e}sert, Flagg, Gill, Inglis, Kirk, Knutson,
  Kreidberg, L{\'o}pez~Morales, Mansfield, Moran, Murray, Nixon, {Petit dit de
  la Roche}, Rackham, Schlawin, Sing, Wakeford, Wallack, Wheatley, Zieba,
  Aggarwal, Barstow, Bell, Blecic, Caceres, Crouzet, Cubillos, Daylan, {de
  Val-Borro}, Decin, Fortney, Gibson, Heng, Hu, Kempton, Lagage, Lothringer,
  {Lustig-Yaeger}, Mancini, Mayne, Mayorga, Molaverdikhani, Nasedkin, Ohno,
  Parmentier, Powell, Redfield, Roy, Taylor, \& Zhang}]{Carter2024}
Carter, A.~L., May, E.~M., Espinoza, N., {et~al.} 2024, Nature Astronomy, 1,
  \dodoi{10.1038/s41550-024-02292-x}

\bibitem[{Charbonneau {et~al.}(2002)Charbonneau, Brown, Noyes, \&
  Gilliland}]{Charbonneau2002}
Charbonneau, D., Brown, T.~M., Noyes, R.~W., \& Gilliland, R.~L. 2002, The
  Astrophysical Journal, 568, 377, \dodoi{10.1086/338770}

\bibitem[{Cont {et~al.}(2024)Cont, Nortmann, Yan, Lesjak, Czesla, Lavail,
  Reiners, Piskunov, Hatzes, {Boldt-Christmas}, Kochukhov, Marquart, Nagel,
  Rains, Rengel, Seemann, \& Shulyak}]{Cont2024}
Cont, D., Nortmann, L., Yan, F., {et~al.} 2024, Astronomy \& Astrophysics, 688,
  A206, \dodoi{10.1051/0004-6361/202450064}

\bibitem[{Cooper \& Showman(2005)}]{Cooper2005}
Cooper, C.~S., \& Showman, A.~P. 2005, The Astrophysical Journal, 629, L45,
  \dodoi{10.1086/444354}

\bibitem[{{Dobbs-Dixon} {et~al.}(2012){Dobbs-Dixon}, Agol, \&
  Burrows}]{Dobbs-Dixon2012}
{Dobbs-Dixon}, I., Agol, E., \& Burrows, A. 2012, The Astrophysical Journal,
  751, 87, \dodoi{10.1088/0004-637X/751/2/87}

\bibitem[{Drummond {et~al.}(2020)Drummond, H{\'e}brard, Mayne, Venot, Ridgway,
  Changeat, Tsai, Manners, Tremblin, Abraham, Sing, \&
  Koh{\'a}ry}]{Drummond2020}
Drummond, B., H{\'e}brard, E., Mayne, N.~J., {et~al.} 2020, Astronomy and
  Astrophysics, \dodoi{10.1051/0004-6361/201937153}

\bibitem[{Ehrenreich {et~al.}(2020)Ehrenreich, Lovis, Allart, Zapatero~Osorio,
  Pepe, Cristiani, Rebolo, Santos, Borsa, Demangeon, Dumusque,
  Gonz{\'a}lez~Hern{\'a}ndez, {Casasayas-Barris}, S{\'e}gransan, Sousa, Abreu,
  Adibekyan, Affolter, Allende~Prieto, Alibert, Aliverti, Alves, Amate, Avila,
  Baldini, Bandy, Benz, Bianco, Bolmont, Bouchy, Bourrier, Broeg, Cabral,
  Calderone, Pall{\'e}, Cegla, Cirami, Coelho, Conconi, Coretti, Cumani,
  Cupani, Dekker, Delabre, Deiries, D'Odorico, Di~Marcantonio, Figueira,
  Fragoso, Genolet, Genoni, G{\'e}nova~Santos, Hara, Hughes, Iwert, Kerber,
  Knudstrup, Landoni, Lavie, Lizon, Lendl, Lo~Curto, Maire, Manescau, Martins,
  M{\'e}gevand, Mehner, Micela, Modigliani, Molaro, Monteiro, Monteiro,
  Moschetti, M{\"u}ller, Nunes, Oggioni, Oliveira, Pariani, Pasquini, Poretti,
  Rasilla, Redaelli, Riva, Santana~Tschudi, Santin, Santos, Segovia~Milla,
  Seidel, Sosnowska, Sozzetti, Span{\`o}, Su{\'a}rez~Mascare{\~n}o, Tabernero,
  Tenegi, Udry, Zanutta, \& Zerbi}]{Ehrenreich2020}
Ehrenreich, D., Lovis, C., Allart, R., {et~al.} 2020, Nature, 580, 597,
  \dodoi{10.1038/s41586-020-2107-1}

\bibitem[{Espinoza \& Jones(2021)}]{Espinoza2021}
Espinoza, N., \& Jones, K. 2021, The Astronomical Journal, 162, 165,
  \dodoi{10.3847/1538-3881/ac134d}

\bibitem[{Espinoza {et~al.}(2024)Espinoza, Steinrueck, Kirk, MacDonald, Savel,
  Arnold, Kempton, Murphy, Carone, Zamyatina, Lewis, Samra, Kiefer, Rauscher,
  Christie, Mayne, Helling, Rustamkulov, Parmentier, May, Carter, Zhang,
  {L{\'o}pez-Morales}, Allen, Blecic, Decin, Mancini, Molaverdikhani, Rackham,
  Palle, Tsai, Ahrer, Bean, Crossfield, Haegele, H{\'e}brard, Kreidberg,
  Powell, Schneider, Welbanks, Wheatley, Brahm, \& Crouzet}]{Espinoza2024}
Espinoza, N., Steinrueck, M.~E., Kirk, J., {et~al.} 2024, Nature,
  \dodoi{10.1038/s41586-024-07768-4}

\bibitem[{Faedi {et~al.}(2011)Faedi, Barros, Anderson, Brown, Cameron,
  Pollacco, Boisse, Hebrard, Lendl, Lister, {et~al.}}]{faedi2011wasp}
Faedi, F., Barros, S.~C., Anderson, D.~R., {et~al.} 2011, Astronomy \&
  Astrophysics, 531, A40

\bibitem[{Fairman {et~al.}(2024)Fairman, Wakeford, \& MacDonald}]{Fairman2024}
Fairman, C., Wakeford, H.~R., \& MacDonald, R.~J. 2024, The Astronomical
  Journal, 167, 240.
\newblock \doarXiv{2403.07801}

\bibitem[{Feinstein {et~al.}(2023)Feinstein, Radica, Welbanks, Murray, Ohno,
  Coulombe, Espinoza, Bean, Teske, Benneke, {et~al.}}]{feinstein2023early}
Feinstein, A.~D., Radica, M., Welbanks, L., {et~al.} 2023, Nature, 614, 670

\bibitem[{Fischer {et~al.}(2016)Fischer, Knutson, Sing, Henry, Williamson,
  Fortney, Burrows, Kataria, Nikolov, Showman, {et~al.}}]{fischer2016hst}
Fischer, P.~D., Knutson, H.~A., Sing, D.~K., {et~al.} 2016, The Astrophysical
  Journal, 827, 19

\bibitem[{Fortney {et~al.}(2010)Fortney, Shabram, Showman, Lian, Freedman,
  Marley, \& Lewis}]{Fortney2010}
Fortney, J.~J., Shabram, M., Showman, A.~P., {et~al.} 2010, The Astrophysical
  Journal, 709, 1396, \dodoi{10.1088/0004-637X/709/2/1396}

\bibitem[{{Helling} {et~al.}(2023){Helling}, {Samra}, {Lewis}, {Calder},
  {Hirst}, {Woitke}, {Baeyens}, {Carone}, {Herbort}, \&
  {Chubb}}]{2023A&A...671A.122H}
{Helling}, C., {Samra}, D., {Lewis}, D., {et~al.} 2023, \aap, 671, A122,
  \dodoi{10.1051/0004-6361/202243956}

\bibitem[{Holton(1992)}]{Holton1992}
Holton, J.~R. 1992, An Introduction to Dynamic Meteorology

\bibitem[{Ji {et~al.}(2024)Ji, Li, Zhang, Li, Fang, Wang, Deng, Chen, Li, Dong,
  Li, Gao, \& Xian}]{2024ChJSS..44..193J}
Ji, J., Li, H., Zhang, J., {et~al.} 2024, Chinese Journal of Space Science, 44,
  193, \dodoi{10.11728/cjss2024.02.yg03}

\bibitem[{Ji {et~al.}(2022)Ji, Li, Zhang, Fang, Li, Wang, Cao, Deng, Li, Xian,
  Gao, Zhang, Li, Liu, Qi, Jin, Liu, Chen, Li, Dong, \& Zhu}]{Ji2022}
Ji, J.-H., Li, H.-T., Zhang, J.-B., {et~al.} 2022, Research in Astronomy and
  Astrophysics, 22, 072003, \dodoi{10.1088/1674-4527/ac77e4}

\bibitem[{Jones \& Espinoza(2022)}]{Jones2022}
Jones, K., \& Espinoza, N. 2022, Journal of Open Source Software, 7, 2382,
  \dodoi{10.21105/joss.02382}

\bibitem[{{JWST Transiting Exoplanet Community Early Release Science Team}
  {et~al.}(2023){JWST Transiting Exoplanet Community Early Release Science
  Team}, {Ahrer}, {Alderson}, {Batalha}, {Batalha}, {Bean}, {Beatty}, {Bell},
  {Benneke}, {Berta-Thompson}, {Carter}, {Crossfield}, {Espinoza}, {Feinstein},
  {Fortney}, {Gibson}, {Goyal}, {Kempton}, {Kirk}, {Kreidberg},
  {L{\'o}pez-Morales}, {Line}, {Lothringer}, {Moran}, {Mukherjee}, {Ohno},
  {Parmentier}, {Piaulet}, {Rustamkulov}, {Schlawin}, {Sing}, {Stevenson},
  {Wakeford}, {Allen}, {Birkmann}, {Brande}, {Crouzet}, {Cubillos}, {Damiano},
  {D{\'e}sert}, {Gao}, {Harrington}, {Hu}, {Kendrew}, {Knutson}, {Lagage},
  {Leconte}, {Lendl}, {MacDonald}, {May}, {Miguel}, {Molaverdikhani}, {Moses},
  {Murray}, {Nehring}, {Nikolov}, {Petit dit de la Roche}, {Radica}, {Roy},
  {Stassun}, {Taylor}, {Waalkes}, {Wachiraphan}, {Welbanks}, {Wheatley},
  {Aggarwal}, {Alam}, {Banerjee}, {Barstow}, {Blecic}, {Casewell}, {Changeat},
  {Chubb}, {Col{\'o}n}, {Coulombe}, {Daylan}, {de Val-Borro}, {Decin}, {Dos
  Santos}, {Flagg}, {France}, {Fu}, {Garc{\'\i}a Mu{\~n}oz}, {Gizis},
  {Glidden}, {Grant}, {Heng}, {Henning}, {Hong}, {Inglis}, {Iro}, {Kataria},
  {Komacek}, {Krick}, {Lee}, {Lewis}, {Lillo-Box}, {Lustig-Yaeger}, {Mancini},
  {Mandell}, {Mansfield}, {Marley}, {Mikal-Evans}, {Morello}, {Nixon}, {Ortiz
  Ceballos}, {Piette}, {Powell}, {Rackham}, {Ramos-Rosado}, {Rauscher},
  {Redfield}, {Rogers}, {Roman}, {Roudier}, {Scarsdale}, {Shkolnik},
  {Southworth}, {Spake}, {Steinrueck}, {Tan}, {Teske}, {Tremblin}, {Tsai},
  {Tucker}, {Turner}, {Valenti}, {Venot}, {Waldmann}, {Wallack}, {Zhang}, \&
  {Zieba}}]{2023Natur.614..649J}
{JWST Transiting Exoplanet Community Early Release Science Team}, {Ahrer},
  E.-M., {Alderson}, L., {et~al.} 2023, \nat, 614, 649,
  \dodoi{10.1038/s41586-022-05269-w}

\bibitem[{Karman {et~al.}(2019)Karman, Gordon, Van Der~Avoird, Baranov, Boulet,
  Drouin, Groenenboom, Gustafsson, Hartmann, Kurucz, Rothman, Sun, Sung,
  Thalman, Tran, Wishnow, Wordsworth, Vigasin, Volkamer, \& Van
  Der~Zande}]{Karman2019}
Karman, T., Gordon, I.~E., Van Der~Avoird, A., {et~al.} 2019, Icarus, 328, 160,
  \dodoi{10.1016/j.icarus.2019.02.034}

\bibitem[{Kawashima \& Min(2021)}]{Kawashima2021}
Kawashima, Y., \& Min, M. 2021, Astronomy \& Astrophysics,
  \dodoi{10.1051/0004-6361/202141548}

\bibitem[{Kempton {et~al.}(2017)Kempton, Bean, \& Parmentier}]{Kempton2017}
Kempton, E. M.-R., Bean, J.~L., \& Parmentier, V. 2017, The Astrophysical
  Journal Letters, 845, L20, \dodoi{10.3847/2041-8213/aa84ac}

\bibitem[{Kirk {et~al.}(2019)Kirk, {L{\'o}pez-Morales}, Wheatley, Weaver,
  Skillen, Louden, McCormac, \& Espinoza}]{kirk2019lrg}
Kirk, J., {L{\'o}pez-Morales}, M., Wheatley, P.~J., {et~al.} 2019, The
  Astronomical Journal, 158, 144

\bibitem[{{Knutson} {et~al.}(2007){Knutson}, {Charbonneau}, {Allen}, {Fortney},
  {Agol}, {Cowan}, {Showman}, {Cooper}, \& {Megeath}}]{2007Natur.447..183K}
{Knutson}, H.~A., {Charbonneau}, D., {Allen}, L.~E., {et~al.} 2007, \nat, 447,
  183, \dodoi{10.1038/nature05782}

\bibitem[{Landgren \& Nadeau(2022)}]{Landgren2022}
Landgren, E., \& Nadeau, A. 2022, Journal of Open Source Software, 7, 4872,
  \dodoi{10.21105/joss.04872}

\bibitem[{Landgren {et~al.}(2023)Landgren, Nadeau, Lewis, Kataria, \&
  Hitchcock}]{Landgren2023}
Landgren, E., Nadeau, A., Lewis, N., Kataria, T., \& Hitchcock, P. 2023, The
  Planetary Science Journal, 4, 106, \dodoi{10.3847/PSJ/acd551}

\bibitem[{Li {et~al.}(2023)Li, Chen, Zhao, \& Wang}]{Li2023}
Li, X.-K., Chen, G., Zhao, H.-B., \& Wang, H.-C. 2023, Research in Astronomy
  and Astrophysics, 23, 025018, \dodoi{10.1088/1674-4527/acae71}

\bibitem[{Line \& Parmentier(2016)}]{Line2016}
Line, M.~R., \& Parmentier, V. 2016, The Astrophysical Journal, 820, 78,
  \dodoi{10.3847/0004-637X/820/1/78}

\bibitem[{{MacDonald} \& {Lewis}(2022)}]{2022ApJ...929...20M}
{MacDonald}, R.~J., \& {Lewis}, N.~K. 2022, \apj, 929, 20,
  \dodoi{10.3847/1538-4357/ac47fe}

\bibitem[{{MacDonald} \& {Madhusudhan}(2017)}]{2017MNRAS.469.1979M}
{MacDonald}, R.~J., \& {Madhusudhan}, N. 2017, \mnras, 469, 1979,
  \dodoi{10.1093/mnras/stx804}

\bibitem[{Mayne {et~al.}(2013)Mayne, Baraffe, Acreman, Smith, Browning,
  Amundsen, Wood, Thuburn, \& Jackson}]{Mayne2013}
Mayne, N.~J., Baraffe, I., Acreman, D.~M., {et~al.} 2013, Astronomy and
  Astrophysics, \dodoi{10.1051/0004-6361/201322174}

\bibitem[{Mayne {et~al.}(2017)Mayne, Debras, Baraffe, Thuburn, Amundsen,
  Acreman, Smith, Browning, Manners, \& Wood}]{Mayne2017}
Mayne, N.~J., Debras, F., Baraffe, I., {et~al.} 2017, Astronomy and
  Astrophysics, \dodoi{10.1051/0004-6361/201730465}

\bibitem[{Menou \& Rauscher(2009)}]{Menou2009}
Menou, K., \& Rauscher, E. 2009, The Astrophysical Journal,
  \dodoi{10.1088/0004-637x/700/1/887}

\bibitem[{Murphy {et~al.}(2024)Murphy, Beatty, Schlawin, Bell, Line, Greene,
  Parmentier, Rauscher, Welbanks, Fortney, \& Rieke}]{Murphy2024}
Murphy, M.~M., Beatty, T.~G., Schlawin, E., {et~al.} 2024, Nature Astronomy.
\newblock \doarXiv{2406.09863}

\bibitem[{Nikolov {et~al.}(2016)Nikolov, Sing, Gibson, Fortney, Evans, Barstow,
  Kataria, \& Wilson}]{Nikolov2016}
Nikolov, N., Sing, D.~K., Gibson, N.~P., {et~al.} 2016, The Astrophysical
  Journal, 832, 191, \dodoi{10.3847/0004-637X/832/2/191}

\bibitem[{Nortmann {et~al.}(2024)Nortmann, Lesjak, Yan, Cont, Czesla, Lavail,
  Rains, Nagel, {Boldt-Christmas}, Hatzes, Reiners, Piskunov, Kochukhov,
  Heiter, Shulyak, Rengel, \& Seemann}]{Nortmann2024}
Nortmann, L., Lesjak, F., Yan, F., {et~al.} 2024,  arXiv.
\newblock \doarXiv{2404.12363}

\bibitem[{Parmentier {et~al.}(2021)Parmentier, Showman, \&
  Fortney}]{2021MNRAS.501...78P}
Parmentier, V., Showman, A.~P., \& Fortney, J.~J. 2021, 501, 78,
  \dodoi{10.1093/mnras/staa3418}

\bibitem[{{Perez-Becker} \& Showman(2013)}]{Perez-Becker2013}
{Perez-Becker}, D., \& Showman, A.~P. 2013, The Astrophysical Journal, 776,
  134, \dodoi{10.1088/0004-637X/776/2/134}

\bibitem[{Pinhas {et~al.}(2019)Pinhas, Madhusudhan, Gandhi, \&
  MacDonald}]{pinhas2019h2o}
Pinhas, A., Madhusudhan, N., Gandhi, S., \& MacDonald, R. 2019, Monthly Notices
  of the Royal Astronomical Society, 482, 1485

\bibitem[{Pinhas {et~al.}(2018)Pinhas, Rackham, Madhusudhan, \&
  Apai}]{pinhas2018retrieval}
Pinhas, A., Rackham, B.~V., Madhusudhan, N., \& Apai, D. 2018, Monthly Notices
  of the Royal Astronomical Society, 480, 5314

\bibitem[{Powell {et~al.}(2019)Powell, Louden, Kreidberg, Zhang, Gao, \&
  Parmentier}]{Powell2019}
Powell, D., Louden, T., Kreidberg, L., {et~al.} 2019, The Astrophysical
  Journal, 887, 170, \dodoi{10.3847/1538-4357/ab55d9}

\bibitem[{Powell \& Zhang(2024)}]{Powell2024a}
Powell, D., \& Zhang, X. 2024, The Astrophysical Journal, 969, 5,
  \dodoi{10.3847/1538-4357/ad3de4}

\bibitem[{{Powell} {et~al.}(2024){Powell}, {Feinstein}, {Lee}, {Zhang}, {Tsai},
  {Taylor}, {Kirk}, {Bell}, {Barstow}, {Gao}, {Bean}, {Blecic}, {Chubb},
  {Crossfield}, {Jordan}, {Kitzmann}, {Moran}, {Morello}, {Moses}, {Welbanks},
  {Yang}, {Zhang}, {Ahrer}, {Bello-Arufe}, {Brande}, {Casewell}, {Crouzet},
  {Cubillos}, {Demory}, {Dyrek}, {Flagg}, {Hu}, {Inglis}, {Jones}, {Kreidberg},
  {L{\'o}pez-Morales}, {Lagage}, {Meier Vald{\'e}s}, {Miguel}, {Parmentier},
  {Piette}, {Rackham}, {Radica}, {Redfield}, {Stevenson}, {Wakeford},
  {Aggarwal}, {Alam}, {Batalha}, {Batalha}, {Benneke}, {Berta-Thompson},
  {Brady}, {Caceres}, {Carter}, {D{\'e}sert}, {Harrington}, {Iro}, {Line},
  {Lothringer}, {MacDonald}, {Mancini}, {Molaverdikhani}, {Mukherjee}, {Nixon},
  {Oza}, {Palle}, {Rustamkulov}, {Sing}, {Steinrueck}, {Venot}, {Wheatley}, \&
  {Yurchenko}}]{2024Natur.626..979P}
{Powell}, D., {Feinstein}, A.~D., {Lee}, E. K.~H., {et~al.} 2024, \nat, 626,
  979, \dodoi{10.1038/s41586-024-07040-9}

\bibitem[{Rauscher \& Menou(2012)}]{Rauscher2012}
Rauscher, E., \& Menou, K. 2012, The Astrophysical Journal,
  \dodoi{10.1088/0004-637x/750/2/96}

\bibitem[{Richard {et~al.}(2012)Richard, Gordon, Rothman, Abel, Frommhold,
  Gustafsson, Hartmann, Hermans, Lafferty, Orton, Smith, \& Tran}]{Richard2012}
Richard, C., Gordon, I., Rothman, L., {et~al.} 2012, Journal of Quantitative
  Spectroscopy and Radiative Transfer, 113, 1276,
  \dodoi{10.1016/j.jqsrt.2011.11.004}

\bibitem[{Roman {et~al.}(2021)Roman, Kempton, Rauscher, Harada, Bean, \&
  Stevenson}]{2021ApJ...908..101R}
Roman, M.~T., Kempton, E. M.~R., Rauscher, E., {et~al.} 2021, 908, 101,
  \dodoi{10.3847/1538-4357/abd549}

\bibitem[{Roth {et~al.}(2024)Roth, Parmentier, \& Hammond}]{Roth2024}
Roth, A., Parmentier, V., \& Hammond, M. 2024, Monthly Notices of the Royal
  Astronomical Society, stae984, \dodoi{10.1093/mnras/stae984}

\bibitem[{Rustamkulov {et~al.}(2023)Rustamkulov, Sing, Mukherjee, May, Kirk,
  Schlawin, Line, Piaulet, Carter, Batalha, Goyal, {L{\'o}pez-Morales},
  Lothringer, MacDonald, Moran, Stevenson, Wakeford, Espinoza, Bean, Batalha,
  Benneke, {Berta-Thompson}, Crossfield, Gao, Kreidberg, Powell, Cubillos,
  Gibson, Leconte, Molaverdikhani, Nikolov, Parmentier, Roy, Taylor, Turner,
  Wheatley, Aggarwal, Ahrer, Alam, Alderson, Allen, Banerjee, Barat, Barrado,
  Barstow, Bell, Blecic, Brande, Casewell, Changeat, Chubb, Crouzet, Daylan,
  Decin, D{\'e}sert, {Mikal-Evans}, Feinstein, Flagg, Fortney, Harrington,
  Heng, Hong, Hu, Iro, Kataria, Kempton, Krick, Lendl, {Lillo-Box}, Louca,
  {Lustig-Yaeger}, Mancini, Mansfield, Mayne, Miguel, Morello, Ohno, Palle,
  {Petit dit de la Roche}, Rackham, Radica, {Ramos-Rosado}, Redfield, Rogers,
  Shkolnik, Southworth, Teske, Tremblin, Tucker, Venot, Waalkes, Welbanks,
  Zhang, \& Zieba}]{Rustamkulov2023}
Rustamkulov, Z., Sing, D.~K., Mukherjee, S., {et~al.} 2023, Nature, 1,
  \dodoi{10.1038/s41586-022-05677-y}

\bibitem[{Showman {et~al.}(2009)Showman, Fortney, Lian, Marley, Freedman,
  Knutson, \& Charbonneau}]{Showman2009}
Showman, A.~P., Fortney, J.~J., Lian, Y., {et~al.} 2009, The Astrophysical
  Journal, 699, 564, \dodoi{10.1088/0004-637X/699/1/564}

\bibitem[{Showman \& Guillot(2002)}]{Showman2002}
Showman, A.~P., \& Guillot, T. 2002, Astronomy \& Astrophysics, 385, 166,
  \dodoi{10.1051/0004-6361:20020101}

\bibitem[{Sing {et~al.}(2016)Sing, Fortney, Nikolov, Wakeford, Kataria, Evans,
  Aigrain, Ballester, Burrows, Deming, Desert, Gibson, Henry, Huitson, Knutson,
  {des Etangs}, Pont, Showman, {Vidal-Madjar}, Williamson, \&
  Wilson}]{Sing2016}
Sing, D.~K., Fortney, J.~J., Nikolov, N., {et~al.} 2016, NATURE, 529, 59,
  \dodoi{10.1038/nature16068}

\bibitem[{{Stevenson} {et~al.}(2014){Stevenson}, {D{\'e}sert}, {Line}, {Bean},
  {Fortney}, {Showman}, {Kataria}, {Kreidberg}, {McCullough}, {Henry},
  {Charbonneau}, {Burrows}, {Seager}, {Madhusudhan}, {Williamson}, \&
  {Homeier}}]{2014Sci...346..838S}
{Stevenson}, K.~B., {D{\'e}sert}, J.-M., {Line}, M.~R., {et~al.} 2014, Science,
  346, 838, \dodoi{10.1126/science.1256758}

\bibitem[{Stevenson {et~al.}(2016)Stevenson, Lewis, Bean, Beichman, Fraine,
  Kilpatrick, Krick, Lothringer, Mandell, Valenti, Agol, Angerhausen, Barstow,
  Birkmann, Burrows, Charbonneau, Cowan, Crouzet, Cubillos, Curry, Dalba, {de
  Wit}, Deming, D{\'e}sert, Doyon, Dragomir, Ehrenreich, Fortney,
  Garc{\'\i}a~Mu{\~n}oz, Gibson, Gizis, Greene, Harrington, Heng, Kataria,
  Kempton, Knutson, Kreidberg, Lafreni{\`e}re, Lagage, Line, {Lopez-Morales},
  Madhusudhan, Morley, Rocchetto, Schlawin, Shkolnik, Shporer, Sing, Todorov,
  Tucker, \& Wakeford}]{Stevenson2016}
Stevenson, K.~B., Lewis, N.~K., Bean, J.~L., {et~al.} 2016, Publications of the
  Astronomical Society of the Pacific, 128, 094401,
  \dodoi{10.1088/1538-3873/128/967/094401}

\bibitem[{Tennyson \& Yurchenko(2018)}]{Tennyson2018TheEA}
Tennyson, J., \& Yurchenko, S.~N. 2018, Atoms, 6, 26

\bibitem[{{The Astropy Collaboration} {et~al.}(2013){The Astropy
  Collaboration}, Robitaille, Tollerud, Greenfield, Droettboom, Bray, Aldcroft,
  Davis, Ginsburg, {Price-Whelan}, Kerzendorf, Conley, Crighton, Barbary, Muna,
  Ferguson, Grollier, Parikh, Nair, G{\"u}nther, Deil, Woillez, Conseil,
  Kramer, Turner, Singer, Fox, Weaver, Zabalza, Edwards, Azalee~Bostroem,
  Burke, Casey, Crawford, Dencheva, Ely, Jenness, Labrie, Lim, Pierfederici,
  Pontzen, Ptak, Refsdal, Servillat, \&
  Streicher}]{TheAstropyCollaboration2013}
{The Astropy Collaboration}, Robitaille, T.~P., Tollerud, E.~J., {et~al.} 2013,
  Astronomy \& Astrophysics, 558, A33, \dodoi{10.1051/0004-6361/201322068}

\bibitem[{{The Astropy Collaboration} {et~al.}(2018){The Astropy
  Collaboration}, {Price-Whelan}, Sip{\H o}cz, G{\"u}nther, Lim, Crawford,
  Conseil, Shupe, Craig, Dencheva, Ginsburg, VanderPlas, Bradley,
  {P{\'e}rez-Su{\'a}rez}, {De Val-Borro}, {(Primary Paper Contributors)},
  Aldcroft, Cruz, Robitaille, Tollerud, {(Astropy Coordination Committee)},
  Ardelean, Babej, Bach, Bachetti, Bakanov, Bamford, Barentsen, Barmby,
  Baumbach, Berry, Biscani, Boquien, Bostroem, Bouma, Brammer, Bray,
  Breytenbach, Buddelmeijer, Burke, Calderone, Rodr{\'\i}guez, Cara, Cardoso,
  Cheedella, Copin, Corrales, Crichton, D'Avella, Deil, Depagne, Dietrich,
  Donath, Droettboom, Earl, Erben, Fabbro, Ferreira, Finethy, Fox, Garrison,
  Gibbons, Goldstein, Gommers, Greco, Greenfield, Groener, Grollier, Hagen,
  Hirst, Homeier, Horton, Hosseinzadeh, Hu, Hunkeler, Ivezi{\'c}, Jain,
  Jenness, Kanarek, Kendrew, Kern, Kerzendorf, Khvalko, King, Kirkby, Kulkarni,
  Kumar, Lee, Lenz, Littlefair, Ma, Macleod, Mastropietro, McCully, Montagnac,
  Morris, Mueller, Mumford, Muna, Murphy, Nelson, Nguyen, Ninan, N{\"o}the,
  Ogaz, Oh, Parejko, Parley, Pascual, Patil, Patil, Plunkett, Prochaska,
  Rastogi, Janga, Sabater, Sakurikar, Seifert, Sherbert, {Sherwood-Taylor},
  Shih, Sick, Silbiger, Singanamalla, Singer, Sladen, Sooley, Sornarajah,
  Streicher, Teuben, Thomas, Tremblay, Turner, Terr{\'o}n, Kerkwijk,
  De~La~Vega, Watkins, Weaver, Whitmore, Woillez, Zabalza, \& {(Astropy
  Contributors)}}]{TheAstropyCollaboration2018}
{The Astropy Collaboration}, {Price-Whelan}, A.~M., Sip{\H o}cz, B.~M.,
  {et~al.} 2018, The Astronomical Journal, 156, 123,
  \dodoi{10.3847/1538-3881/aabc4f}

\bibitem[{Trotta(2008)}]{Trotta2008}
Trotta, R. 2008, Contemporary Physics, 49, 71,
  \dodoi{10.1080/00107510802066753}

\bibitem[{Tsai {et~al.}(2023)Tsai, Lee, Powell, Gao, Zhang, Moses, H{\'e}brard,
  Venot, Parmentier, Jordan, Hu, Alam, Alderson, Batalha, Bean, Benneke,
  Bierson, Brady, Carone, Carter, Chubb, Inglis, Leconte, Line,
  {L{\'o}pez-Morales}, Miguel, Molaverdikhani, Rustamkulov, Sing, Stevenson,
  Wakeford, Yang, Aggarwal, Baeyens, Barat, {De Val-Borro}, Daylan, Fortney,
  France, Goyal, Grant, Kirk, Kreidberg, Louca, Moran, Mukherjee, Nasedkin,
  Ohno, Rackham, Redfield, Taylor, Tremblin, Visscher, Wallack, Welbanks,
  Youngblood, Ahrer, Batalha, Behr, {Berta-Thompson}, Blecic, Casewell,
  Crossfield, Crouzet, Cubillos, Decin, D{\'e}sert, Feinstein, Gibson,
  Harrington, Heng, Henning, Kempton, Krick, Lagage, Lendl, Lothringer,
  Mansfield, Mayne, {Mikal-Evans}, Palle, Schlawin, Shorttle, Wheatley, \&
  Yurchenko}]{Tsai2023c}
Tsai, S.-M., Lee, E. K.~H., Powell, D., {et~al.} 2023, Nature, 617, 483,
  \dodoi{10.1038/s41586-023-05902-2}

\bibitem[{Tsiaras {et~al.}(2018)Tsiaras, Waldmann, Zingales, Rocchetto,
  Morello, Damiano, Karpouzas, Tinetti, McKemmish, Tennyson,
  {et~al.}}]{tsiaras2018population}
Tsiaras, A., Waldmann, {\relax IP}., Zingales, T., {et~al.} 2018, The
  Astronomical Journal, 155, 156

\bibitem[{{Vidal-Madjar} {et~al.}(2003){Vidal-Madjar}, {des Etangs},
  D{\'e}sert, Ballester, Ferlet, H{\'e}brard, \& Mayor}]{Vidal-Madjar2003}
{Vidal-Madjar}, A., {des Etangs}, A.~L., D{\'e}sert, J.-M., {et~al.} 2003,
  Nature, 422, 143, \dodoi{10.1038/nature01448}

\bibitem[{Wakeford {et~al.}(2017)Wakeford, Sing, Deming, Lewis, Goyal, Wilson,
  Barstow, Kataria, Drummond, Evans, Carter, Nikolov, Knutson, Ballester, \&
  Mandell}]{Wakeford2017}
Wakeford, H.~R., Sing, D.~K., Deming, D., {et~al.} 2017, The Astronomical
  Journal, 155, 29, \dodoi{10.3847/1538-3881/aa9e4e}

\bibitem[{{Wardenier} {et~al.}(2022){Wardenier}, {Parmentier}, \&
  {Lee}}]{2022MNRAS.510..620W}
{Wardenier}, J.~P., {Parmentier}, V., \& {Lee}, E. K.~H. 2022, \mnras, 510,
  620, \dodoi{10.1093/mnras/stab3432}

\bibitem[{Welbanks \& Madhusudhan(2019)}]{welbanks2019degeneracies}
Welbanks, L., \& Madhusudhan, N. 2019, The Astronomical Journal, 157, 206

\bibitem[{Welbanks \& Madhusudhan(2021)}]{Welbanks2021}
---. 2021, The Astrophysical Journal, 913, 114,
  \dodoi{10.3847/1538-4357/abee94}

\bibitem[{Welbanks \& Madhusudhan(2022)}]{Welbanks2022}
---. 2022, The Astrophysical Journal, 933, 79, \dodoi{10.3847/1538-4357/ac6df1}

\bibitem[{Woitke {et~al.}(2018)Woitke, Helling, Hunter, Millard, Turner,
  Worters, Blecic, \& Stock}]{Woitke2018}
Woitke, P., Helling, {\relax Ch}., Hunter, G.~H., {et~al.} 2018, Astronomy \&
  Astrophysics, 614, A1, \dodoi{10.1051/0004-6361/201732193}

\bibitem[{Yang {et~al.}(2024)Yang, Chen, Wang, \& Yan}]{YangYang2024}
Yang, Y., Chen, G., Wang, S., \& Yan, F. 2024, The Astronomical Journal, 167,
  36, \dodoi{10.3847/1538-3881/ad10a3}

\bibitem[{Zhang {et~al.}(2019)Zhang, Chachan, Kempton, \& Knutson}]{Zhang2019}
Zhang, M., Chachan, Y., Kempton, E. M.-R., \& Knutson, H.~A. 2019, Publications
  of the Astronomical Society of the Pacific, 131, 034501,
  \dodoi{10.1088/1538-3873/aaf5ad}

\bibitem[{Zhang {et~al.}(2020)Zhang, Chachan, Kempton, Knutson, \&
  Chang}]{Zhang2020a}
Zhang, M., Chachan, Y., Kempton, E. M.-R., Knutson, H.~A., \& Chang, W.~H.
  2020, The Astrophysical Journal, 899, 27, \dodoi{10.3847/1538-4357/aba1e6}

\end{thebibliography}
\bibliographystyle{aasjournal}

\end{document}